\documentclass[useAMS,usenatbib]{mn2e}
\usepackage{hyperref}
\usepackage{amssymb}
\usepackage{graphicx}
\usepackage[fleqn]{amsmath}
\usepackage{booktabs}
\usepackage{color,soul}

\begin{document}
\title[Impact of Cosmological Satellites on Stellar Discs]{Impact of Cosmological Satellites on Stellar Discs:\\
  Dissecting One Satellite at a Time}
\author[Hu \& Sijacki]{Shaoran Hu$^1$, Debora Sijacki$^1$\\
$^1$ Institute of Astronomy and Kavli Institute for Cosmology, University of Cambridge, Madingley Road, Cambridge CB3 0HA}

\maketitle

\begin{abstract}
Within the standard hierarchical structure formation scenario, Milky Way-mass
dark matter haloes have hundreds of dark matter subhaloes with mass $\gtrsim
10^8 \, {\rm M_{\odot}}$. Over the lifetime of a galactic disc a fraction of
these may pass close to the central region and interact with the disc. We
extract the properties of subhaloes, such as their mass and trajectories, from
a realistic cosmological simulation to study their potential effect on stellar
discs. We find that massive subhalo impacts can generate disc heating, rings,
bars, warps, lopsidedness as wells as spiral structures in the
disc. Specifically, strong counter-rotating single-armed spiral structures
form each time a massive subhalo passes through the disc. Such single-armed
spirals wind up relatively quickly (over $1-2$~Gyrs) and are generally
followed by co-rotating two-armed spiral structures that both develop and wind
up more slowly.
In our simulations self-gravity in the disc is not very strong
and these spiral structures are found to be kinematic density waves.
We demonstrate that there
is a clear link between each spiral mode in the disc and a given subhalo that
caused it, and by changing the mass of the subhalo we can modulate the
strength of the spirals. Furthermore, we find that the majority of subhaloes interact with the disc
impulsively, such that the strength of spirals generated by subhaloes is 
proportional to the total torque they exert. We
conclude that only a handful of encounters with massive subhaloes is
sufficient for re-generating and sustaining spiral structures in discs over
their entire lifetime.
\end{abstract}


\section{Introduction}
\renewcommand{\thefootnote}{\fnsymbol{footnote}}
\footnotetext[1]{E-mail: shaoran.hu@gmail.com}

The Cold Dark Matter scenario predicts a hierarchical growth of structures in
our Universe, whereby low mass objects form first and more massive structures
are assembled later on in course of merging and accretion. Hence, dark matter
haloes are expected to contain a large number of smaller mass haloes, often
dubbed subhaloes, which are gravitationally bound within. With the aid of high
resolution cosmological simulations, properties of these subhaloes can be
directly studied. For example, the Aquarius simulation \citep{Springel2008}
reports an average mass fraction of $11.2\%$ for subhaloes within $r_{50}$,
where $\sim 300,000$ subhaloes in the highest resolution simulation are
resolved. The abundance function of subhaloes generally agrees well among
different simulations \citep{Bullock2010}, where the abundance of subhaloes
above a given mass threshold is roughly inversely related to their mass
through a power law function, i.e. there are more small mass subhaloes than
large mass subhaloes \citep{Springel2008, Gao2004, Diemand2011}. 

In the hierarchical growth scenario, dark matter haloes accrete infalling
subhaloes continuously. As subhaloes orbit in their host dark matter halo,
their orbits usually decay and they lose mass due to dynamical friction and
tidal stripping. In a semi-analytical study \citet{Taylor2004} found that the
mass loss in a single orbit varies from $25\%$ to $45\%$, depending on the
orbit eccentricity and the concentration of the subhalo \citep[see also
  numerical works by e.g.][]{Boylan2008, Jiang2008}. As a combined result
of both effects, most subhaloes that survive reside in the outer region of the
dark matter halo and are prevalently newly accreted. In fact, \cite{Gao2004} found
that only $8\%$ of the subhalo mass survived from $z = 1$ to $z = 0$.

In our local group observed satellite galaxies are considered to be embedded in dark
matter subhaloes \citep[e.g.][]{Mateo1998, Collins2010}. The observed
velocity curves of local group dwarf irregular (dIrr) and dwarf spheroidal
(dSph) galaxies can be well fit by two components, one following the
visible-light distribution while the other being considered as a dark matter component. When
compared with visible satellites, cosmological simulations often over-predicts 
dark matter subhaloes. This is known as the missing satellites problem
\citep{Klypin1999, Moore1999}. Taking into account the lower luminosity limit
of observations, at least part of the problem can be solved
\citep{Tollerud2008, McConnachie2009, Koposov2015}. It is however also
believed that other factors may be at play. Apart from an alternative dark
matter model, e.g. the Warm Dark Matter model that produces much less small
scale structures \citep[for recent work see e.g.][]{Lovell2014, Bose2017}, one
may solve the missing satellites problem by studying various baryonic
effects. For instance, baryons may alter the central region of the subhaloes
to a core \citep{Navarro1996}, which enhances the tidal stripping effect on
subhaloes \citep{Penarrubia2010}; photoionization background and supernova
feedback can suppress star formation in low mass satellites
\citep{Efstathiou1992, Larson1974, Dekel1986, Sawala2016}; baryonic discs may
destroy the subhaloes through strong tidal effects \citep{DOnghia2010}. Such
baryonic effects reduce the number of dark matter subhaloes and weaken or
prevent formation of visible galaxies in some of the dark matter subhaloes,
making them ``invisible''. One way to probe such ``invisible'' dark matter
subhaloes is through their dynamical interactions with baryonic components,
e.g. streams \citep{Erkal2015}.  

It is moreover expected that ``invisible'' dark
  matter 
  subhaloes may also be detected through their interactions with the stellar
  disc. In fact, stellar discs are  in general found to be very sensitive 
to perturbations, including massive molecular clouds \citep{DOnghia2013} and
large-scale torques \citep{dubinski2009, debuhr2012}. In particular,
\citet{Hu2016} found that a realistic triaxial dark matter halo can lead to
two-armed grand-design spiral structures. Given that subhaloes exist widely in
dark matter haloes, it is necessary to understand their interactions with the
spiral structures.

Gravity-induced spiral structures may develop in strong
self-gravitating discs \citep[see e.g.][]{sellwood2014, Hu2016}. Flocculent,
re-current, multi-armed spiral structures form in this scenario.  However, discs
embedded in a realistic dark matter halo may experience disc thickening
and heating in response to subhaloes, therefore weakening the self-gravitating effect. The
vertical scale of the disc in expected to extend by at least $50\%$ to $100\%$, with 
warps also developing in the disc \citep{Velazquez1999, Kazantzidis2009,
  Weinberg1998}. Moreover, the velocity dispersion of stars in the vertical direction is
found to increase when satellites interact with the disc
\citep{Moetazedian2016, Gomez2017}, which leads to an increase in Toomre's $Q$
parameter and can stabilise the disc against self-gravitating modes. The
mass of the subhalo plays a key role in these interactions. \cite{Moetazedian2016}
found that only relatively massive subhaloes ($M>10^9M_\odot$) contribute
towards vertical heating, while \cite{Grand2016} and \cite{Gomez2017} found that
the dominating effect comes from a few satellites with $M>10^{10}M_\odot$.

Previous studies with single test subhaloes have shown that massive subhaloes
impacting with discs can generate grand-design spiral structures \citep[but
  see also][]{Dubinski2008}. \citet{Purcell2011} found that subhaloes of mass $\sim 10^{10.5}\,\mathrm{M_\odot}$ to
$10^{11}\,\mathrm{M_\odot}$ passing as close as $30\,\mathrm{kpc}$ from the disc centre are able
to generate realistic two-armed grand-design spiral structures.
\citet{Pettitt2016} found that the mass limit can be pushed as low as
$10^9\,\mathrm{M_\odot}$ with a closer impact point of $12\,\mathrm{kpc}$, where very weak two-armed
spiral structures can be seen from the Fourier analysis. They also demonstrated
that the spiral response in gaseous
and stellar components are very similar.  

In this work we aim to study the
influence of subhaloes on stellar discs with a more realistic setup to address
the following problems: {\it a)} how subhaloes with realistic properties and trajectories extracted from
cosmological simulations 
interact with the stellar disc, {\it b)} how multiple
subhaloes impacting the disc in succession influence each other, and
{\it c)} how the strength of the spiral structures depends on the properties of the
subhaloes. 

In Section~\ref{sec:method}, we introduce our simulation setup. We present the
dark matter properties of our main simulation in Section~\ref{sec:sssrs},
where realistic subhaloes impacting the central disc are identified. Disc
heating due to subhaloes is discussed in
Section~\ref{sec:disc-heating}. Non-axisymmetric modes including the spiral 
structures that develop are studied in Section~\ref{sec:modes}. The kinematic
properties of these modes are then studied in Section~\ref{sec:powers}. We
then demonstrate the link between the subhalo properties and spiral modes in
Sections~\ref{sec:eimp}, \ref{sec:flyby} and \ref{sec:tidally-driven}.  Finally, in
Section~\ref{sec:conclusion} we present our conclusions.

\section{Method}
\label{sec:method}

\subsection{The Numerical Approach}
\label{sec:numerical}

All simulations in this paper are performed with \textsc{gadget-3}, an updated
version of \textsc{gadget-2} \citep{Springel2005b}. \textsc{gadget-3} is an
$N$-body/SPH code, where different simulated components, such as dark matter and
stars, are represented with particles. The gravitational interaction of the
particles are calculated with the TreePM method.

We model our static dark matter halo based on the Aq-A-4 dark matter halo from
the Aquarius simulation \citep{Springel2008}. The Aquarius simulations is a
dark matter-only cosmological simulation aimed to reproduce Milky Way-sized
haloes. At $z=0$ the Aq-A-4 halo has the virial mass of
$M_{200}=1.84\times 10^{12} \mathrm{M_\odot}$ within the virial radius
$r_{200}=246\,\mathrm{kpc}$, consisting of $1.85\times 10^7$ dark matter
particles, each with a mass of $m_\mathrm{DM}=3.93\times 10^5\,\mathrm{M_\odot}$ and a
softening length of $\epsilon=342\,\mathrm{pc}$. Detailed description of how
we model the dark matter halo from the Aquarius simulation is included in 
Section~\ref{sec:phase1}. 

The stellar disc is initially modeled with an exponential surface density
profile and a vertical isothermal sheet profile following
\citet{springel2005}, i.e.
\begin{equation}
  \label{eq:dm}
  \rho_\mathrm{*}(R,z)=\frac{M_\mathrm{*}}{4\pi z_{\rm 0} R_\mathrm{S}^2} \mathrm{sech}^2\left(\frac{z}{z_0}\right)\exp\left(-\frac{R}{R_\mathrm{S}}\right)\,,
\end{equation}
where $M_\mathrm{*} = 9.5\times 10^9\,\mathrm{M_\odot}$ is the total mass of the disc,
$R_{\rm S} = 3.13\,\mathrm{kpc}$ is the scale length of the disc and $z_{\rm
  0} = 0.1\,{R_{\rm S}}$ is the scale height of the disc. We choose these
parameters so that they match with the high-$Q$ disc in \citet{Hu2016}. As
shown in \citet{Hu2016}, high-$Q$ disc responds to external torques, but does
not form any noticeable self-gravity-induced spiral structures due to its low
surface density, thus enabling us to focus on the kinematic properties. Also,
as found in \citet{Hu2016}, stellar discs, setup to be in equilibrium within a
spherical halo, develop two-armed grand-design spirals when placed in a
triaxial halo directly. To avoid this and study the effect of subhaloes, we
grow the disc in an adiabatically changing dark matter halo, explained in
detail in Section \ref{sec:phase2}. 

To account for baryonic effects (i.e. the influence of baryons on halo shape)
but exclude any unwanted perturbations due to the discreteness noise, we setup
our initial conditions in three steps as follows:
\begin{itemize}
\item \textbf{Phase-1:} We restart the cosmological simulation of the Aq-A-4
  halo from $z = 1.3$ with adiabatically grown static,
  stellar disc potential to obtain the dark matter halo profile for the final
  simulation (Phase-3).  
\item  \textbf{Phase-2:} We then simulate a live stellar disc, with properties
  described above, in an adiabatically changing dark matter halo potential,
  starting from a spherical halo and finishing with the halo shape obtained at
  the end of Phase-1. This step is needed to prepare the initial conditions of
  the stellar disc for the final simulation.
\item \textbf{Phase-3:} In the final step we perform a simulation with a static
  dark matter halo potential (obtained from Phase-1), live dark
  matter subhaloes (directly taken from the Aq-A-4 run) and the live stellar
  disc obtained at the end of Phase-2. With this setup we can isolate the
  impact of subhaloes on the stellar disc. 
\end{itemize}
Further details of the three steps are explained in the following sections.

\subsection{Phase-1: Simulating the Response of the Live Dark Matter Halo to the
Static Stellar Disc Potential}
\label{sec:phase1}

The Aquarius simulation is a dark matter-only simulation hence the effect of
baryons on the dark matter distribution is not taken into account. This is a
reasonable approximation in the outer region of the halo, where the baryonic
matter is not dominant, but may result in a very different inner halo profile
and shape.

To include the effect of baryons, we restart the Aquarius Aq-A-4 simulation
from $z=1.3$, adding a stellar disc potential which is numerically calculated
from the density profile shown in Equation~(\ref{eq:dm}). In the simulation,
we first grow the disc potential  adiabatically from $z=1.3$ to $z=1$.
In the original Aq-A-4 simulation, both the centre and the orientation of the
main halo change over time. We modify the central position and the orientation
of the disc position accordingly, so that the disc is always at
the centre of the halo and aligned with the minor axis of the halo.  We
then keep the disc static from $z=1$ to $z=0$, and study the 
properties of dark matter distribution.

\begin{figure*}
  \centering
  \includegraphics[width=\textwidth]{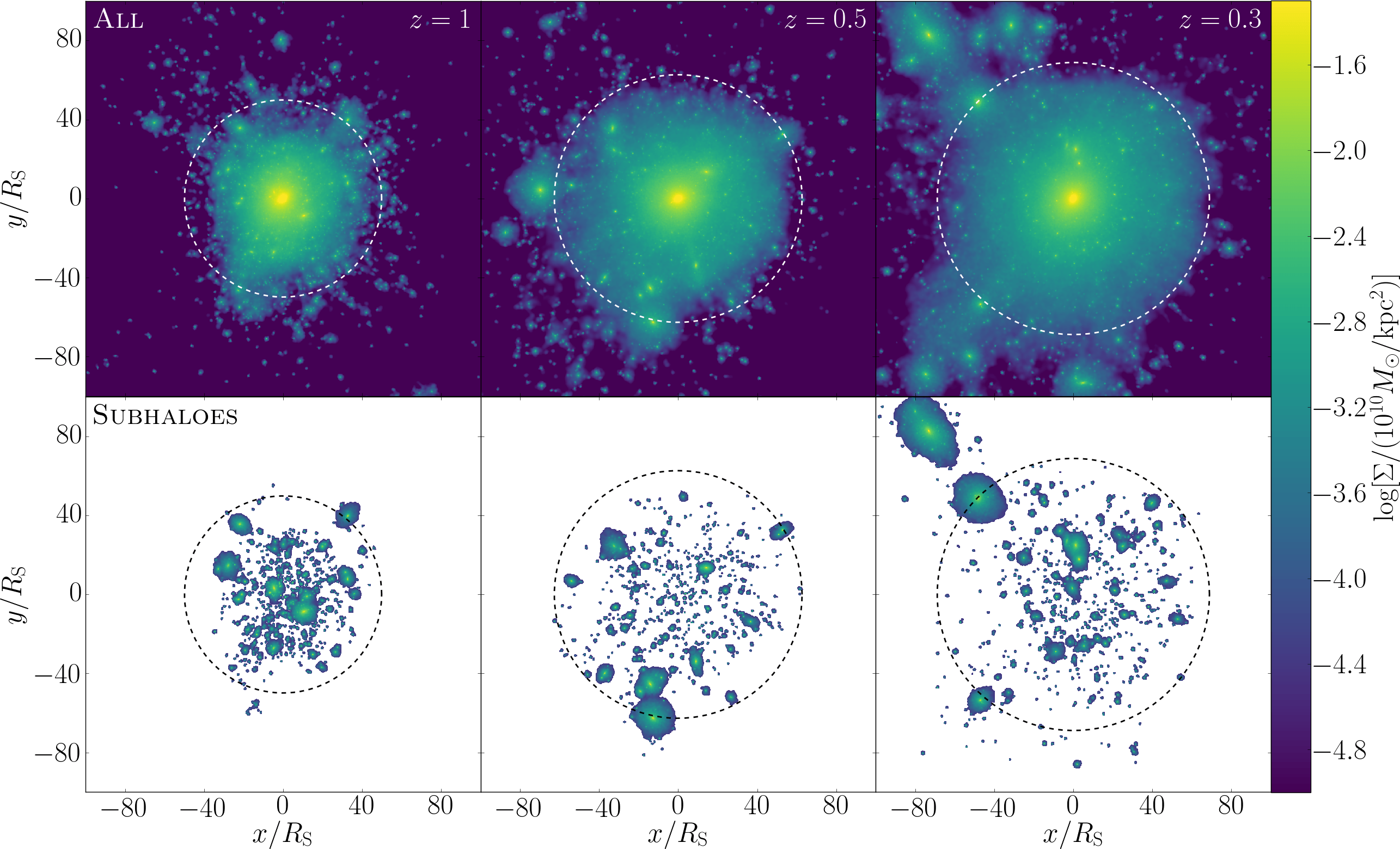}
  \caption{Column density of the dark matter in the Phase-1 simulation
    in the $x-y$ plane at $z = 1$, $z = 0.5$ and $z = 0.3$, from left to right,
    respectively. Top row: the column density of all dark matter
    particles. Bottom row: the column density of subhaloes only. Total mass of the 
    main halo and its subhaloes increases from $1.35\times 10^{11}\,\mathrm{M_\odot}$ to
    $2.29\times 10^{12}\,\mathrm{M_\odot}$ from $z = 1$ to $z = 0$, while the number of
    subhaloes changes from $893$ to $1083$. Dashed circles indicate the virial
    radius at these three epochs. The spatial coordinate is in the unit
      of the scale length of the disc $R_\mathrm{S}=3.13\,\mathrm{kpc}$.}\label{fig:column1}
\end{figure*}

Specifically, the column density distribution of dark matter in the Phase-1
simulation is shown in the top panels of Figure~\ref{fig:column1} at three
different redshifts. The total mass of the halo and its subhaloes increases from
$1.35\times 10^{12}\,\mathrm{M_\odot}$ at $z = 1$ to $2.29\times
10^{12}\,\mathrm{M_\odot}$ at $z = 0$, which is mirrored by the increase in
the physical size of 
the halo as evidenced in the panels. 

We then separate the dark matter component into two groups. We find all
self-bound structures in the
simulation, as identified by the \textsc{subfind} algorithm
\citep{Springel2001, Dolag2009}, at any snapshot between $z = 1$ and $z = 0$. Of
these, all structures that at any time belong to subhaloes, are marked as
subhaloes and are simulated as a ``live'' component in the final simulation
(Phase-3), using their coordinates and velocities at $z = 1$. The bottom
panels of Figure~\ref{fig:column1} show column density distribution of
subhaloes only. All other dark matter particles in the simulation are
considered to be part of a ``smooth'' component (i.e. the difference between
the top and bottom panels) and are represented with an analytic
potential in the Phase-3 simulation (for further details see
Appendix~\ref{sec:ff} and Figure~\ref{fig:1}).

The smooth component can be represented by the sum of a spherical and
a triaxial part. Density profile of the spherical part, $\rho_\mathrm{S}(r)$, is
constructed by spherically averaging the smooth component. We fit it with
three Einasto profiles joined together (see left panel of
Figure~\ref{fig:1} and Appendix~\ref{sec:ff}). Compared to the dark matter
halo profile in the original Aquarius simulation, our profile is more
contracted in the centre due to the disc potential. Our disc
  model has a relatively low mass ($\sim 1\%$) compared to the main halo. This
  is to ensure that the effects due to self-gravity is suppressed, so that we can focus
  on the impact of external perturbations. The
  influence of the disc would be higher with a higher mass model. Following
\citet{bowden2013} and \citet{Hu2016} the triaxial part is represented by two
spherical harmonic functions, aiming to reproduce our desired axis ratio
profile. The axis ratio profile of the smooth component in our simulation is
calculated following a similar method to \citet{Vera-Ciro2011}, where we
iteratively compute the eigenvalues and eigenvectors of the ``reduced''
inertia matrix of an ellipsoid of a certain radius. We find that the axis ratio of the smooth
component significantly fluctuates as a function distance from the centre and
is rather triaxial in the centre (see solid curves in the right panel of
Figure~\ref{fig:1}). This occurs as we only  introduce a stellar disc to
modify the dark matter halo shape and do not consider larger scale
contribution from stars or gas. This would make the disc very elliptical and
unstable. To avoid this issue, we impose a much smoother halo shape variation
with radius (see dashed curves in the right panel of Figure~\ref{fig:1}),
where the outer axis ratios are comparable to those obtained from our Phase-1
simulation.  
 
Having computed $\rho_\mathrm{S}(r)$ and the axis ratios in $x-y$ plane,
$p(r)$, and in $x-z$ plane, $q(r)$, we can calculate the density profile of
the triaxial part as follows
\begin{equation}
  \label{eq:10}
  \rho_\mathrm{T1}(r)=\left( 1-\frac{3q(r)^\alpha}{1+p(r)^\alpha+q(r)^\alpha} \right)\rho_\mathrm{S}(r)
\end{equation}
and
\begin{equation}
  \label{eq:11}
\rho_\mathrm{T2}(r)=\frac{1}{2}\frac{1-p(r)^\alpha}{1+p(r)^\alpha+q^\alpha} \rho_\mathrm{S}(r),
\end{equation}
where $\alpha=-\mathrm{d}\log\rho_\mathrm{S}(r)/\mathrm{d}\log r$ is the slope of
the spherically averaged density profile. We enclose a brief explanation of
these formula in Appendix~\ref{sec:ettp}. Hence, the total density profile of
the halo is 
\begin{equation}
  \label{eq:halo}
  \rho(r,\theta,\varphi)=\rho_\mathrm{S}(r)-\rho_\mathrm{T1}(r)Y_2^0(\theta)+\rho_\mathrm{T2}(r)Y_2^2(\theta,\varphi),
\end{equation}
whose potential can be then integrated numerically using the Green's
function, as shown in Appendix~\ref{sec:cal}.

\subsection{Phase-2: Introducing a Live Disc in a Static, Triaxial Dark
  Matter Halo Potential}
\label{sec:phase2}

The stellar disc is setup initially following Equation~\eqref{eq:dm} and has
$10^6$ stellar particles.  The total mass of the halo inside $5R_\mathrm{S}$ is $\sim 1.6\times 10^{11}M_\odot$,
while the mass of the disc inside $5R_\mathrm{S}$ is $\sim 9\times 10^{9}M_\odot$. The
dynamics of the central disc is therefore dominated by the dark matter halo as expected.
Due to the  high-$Q$ profile of the stellar disc, the swing amplification
is weak. We therefore do not need higher number of particles to prevent transient
gravity-induced spiral structures from forming \citep[for further details see][]{Hu2016}. The stellar disc is initially in equilibrium with a spherical
halo, and as shown in \citet{Hu2016} it will develop strong two-armed spirals
if we put it directly into our smooth triaxial dark matter halo model. To avoid such
structures and focus on the effect of subhaloes alone we have to evolve the
disc inside a dark matter halo that changes adiabatically from spherical to
triaxial. We change the halo in a way similar to Section~3.3 in
\citet{Hu2016}. Namely, we start the simulation with the spherical part of the
halo only, while the triaxial part of the halo grows adiabatically with time,
where the total potential of the halo is
\begin{equation}
  \label{eq:hf}
     \Phi(r,\theta,\phi)=\Phi_{\rm S}(r)+f(t)\Phi_{\rm
     T}(r,\theta,\phi) \,.
 \end{equation}
 Here $\Phi_{\rm S}$ is the static spherical part, $\Phi_{\rm T}$ is the
 triaxial part and the growth factor $f(t)$ changes from $0$ to $1$ following
\begin{equation}
  \label{eq:adia}
  f(t)=(1-\frac{1}{6} \mathrm{e}^{-t/\tau_\mathrm{I}}(1+5\mathrm{e}^{-t/\tau_\mathrm{I}}))^6\,,
\end{equation}
where we have set timescale $\tau_\mathrm{I}=1\,\mathrm{Gyr}$ as in our previous
work. We evolve the system for $5\,\mathrm{Gyr}$ and find no prominent
structures in the disc. The surface density of the final
stellar disc is shown in the top left panel of Figure~\ref{fig:surfdens}.

\subsection{Phase-3: Simulating Live Stellar Disc in Static Triaxial Dark Matter
Halo with Live Subhaloes}
\label{sec:phase3}

With the halo and the disc setup in Phase-1 and Phase-2, we are now able to
perform our final simulation starting from $z = 1$ that studies the influence
of subhaloes on the stellar discs. In this simulation, three components are
included: 
\begin{itemize}
\item A static, triaxial dark matter potential. We
  include it rather than live dark matter particles because it will save
  considerable 
  computational resources and more importantly it will not induce numerical
  perturbations in the disc due to the Poisson noise (note that due to very
  different spatial scales of the disc and the halo, dark matter particle mass
  would need to be considerably larger than the disc particle mass), thus
  making it much more robust to study the influence of subhaloes only.
  Including a static dark matter potential rather than a live
     dark matter halo may however create an artificial torque, as in reality the main
     dark matter halo should gravitationally respond to the impacting
     subhaloes as well. As 
     discussed in detail in Appendix~\ref{sec:fix-main}, we estimate the upper
     limit on this effect and find that it is overall weaker than or at most
     comparable to the genuine effect of
     subhaloes for $m=1$ modes, and negligible for $m=2$ modes. 
  \item Live subhaloes. The
  live subhalo particles are taken directly from the Phase-1 simulation at
  $z=1$. As
  mentioned earlier, they consist of particles that are gravitationally bound to any subhalo
  of our main halo at $z = 1$ or at any later time, based on the
  \textsc{subfind} algorithm. In this way we ensure that
  there is a realistic number of subhaloes during the whole period of the
  Phase-3 simulation. We have in total 2,035,019 live dark matter particles
  included, each having a mass of $m_{\mathrm{DM}} =
  3.93\times 10^5\,\mathrm{M_\odot}$. 
  \item A live disc. The live disc is taken from the Phase-2 simulation at
    $5\,\mathrm{Gyr}$. As explained above, it is in equilibrium with the
    triaxial dark matter potential, and no prominent spiral structures exist in
    the disc. 
\end{itemize}

\section{Results}

\subsection{Properties of Subhaloes that Interact with the Disc}
\label{sec:sssrs}

\begin{figure*}
  \centering
  \includegraphics[width=.95\linewidth]{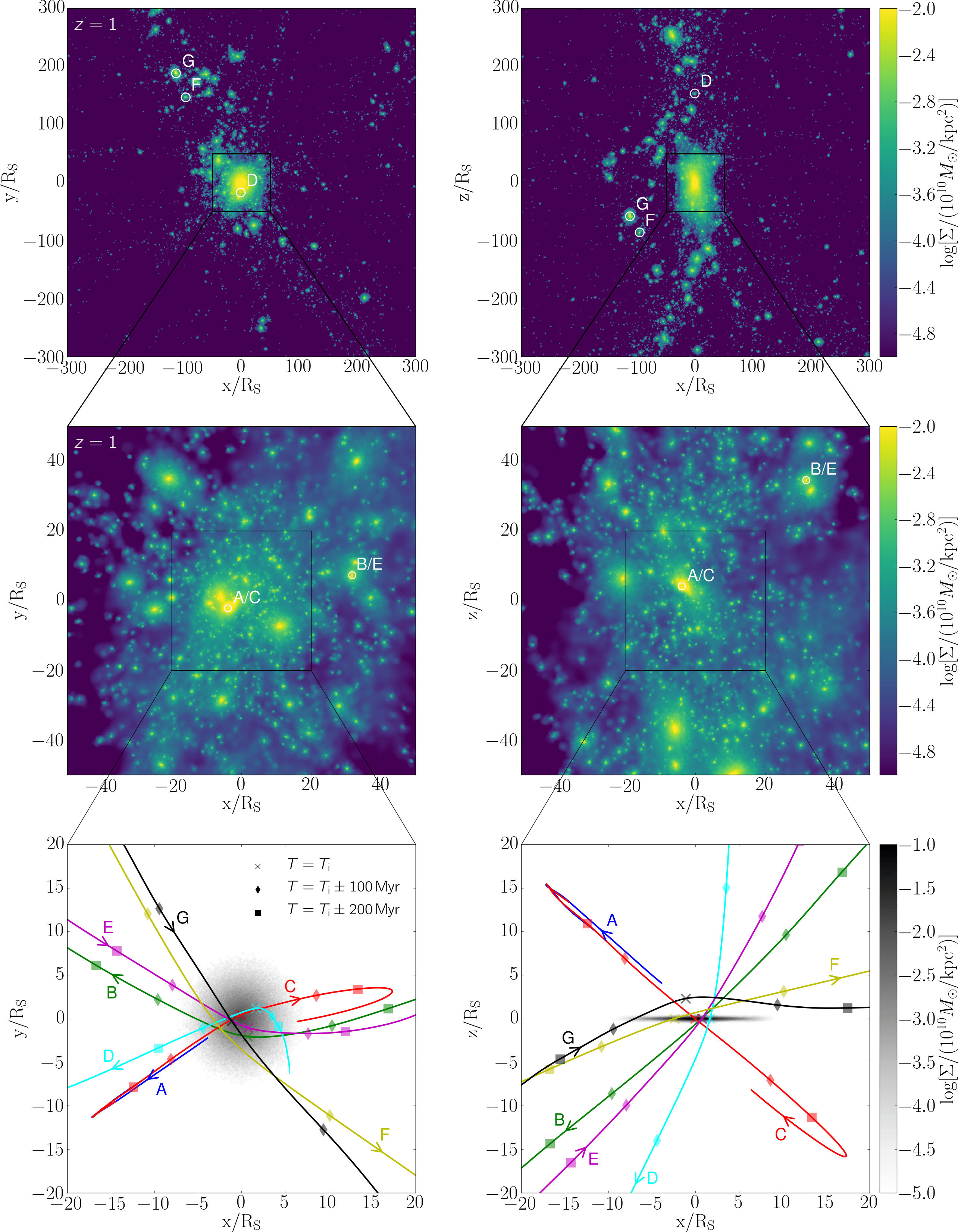}
  \caption{Distribution of dark matter in $x$-$y$ plane (left) and
    $x$-$z$ plane (right). Plots in the top row show the column densities of
    all dark matter particles at $z=1$. Plots in the middle row show the
    column densities of all subhalo particles in the inner halo region at
    $z = 1$. Different letters indicate the initial position of different
    subhaloes which interact with the disc at a later time. Plots in the bottom row
    show the trajectories of the subhaloes that interact with the disc (see
    Table~\ref{tab:1}). In particular, subhaloes A and C are the same subhalo
    hitting the disc at two different redshifts, so are subhaloes B and E.
    The location of subhaloes at the time of impact is marked with
      crosses, while location at $100\,\mathrm{Myr}$ before and after the impact is marked
      with diamonds and at $200\,\mathrm{Myr}$ with squares. The
    column density of the stellar disc is also shown in the bottom row for
    comparison.}\label{fig:traject}
\end{figure*}

\begin{table}
 { \centering
  \caption{Subhaloes that interact with the disc in the Phase-3 simulation.
    Redshift of maximum mass before the impact, $z_{\max{}}$, maximum
    mass before the impact, $M_\mathrm{sub}$, half-mass radius at the
      time of maximum mass before the impact,
    $R_\mathrm{hm}$, redshift of impact,
    $z_\mathrm{i}$, radius of impact, $R_\mathrm{i}$, the angle of incidence,
    $\theta_\mathrm{in}$ and the rotation direction is listed. The half-mass
    radius and the impact radius is shown in units of the scale length of the
    disc, $R_\mathrm{S}=3.13\,\mathrm{kpc}$. The angle of
    incidence is defined as the angle between the trajectory and the normal
    direction of the disc. When $\theta_\mathrm{in} > 90$, the subhalo
    hits the disc from below. For rotation direction, ``R'' stands for
    retrograde, while ``P'' stands for prograde.}\label{tab:1}
  \begin{tabular}{llllllll}
    \toprule
    id & $z_{\max{}}$ & $M_\mathrm{sub}$ & $R_\mathrm{hm}$ & $z_{\mathrm{i}}$  &
    $R_\mathrm{i}$ & $\theta_\mathrm{in}$ & rotation\\
     & &  [$10^{10}\,\mathrm{M_\odot}$] & $/R_\mathrm{S}$ &  & $/R_\mathrm{S}$ & & direction\\
\midrule
    A$^*$ & 1.22 & 1.63   & 1.28   & 1.02  & 1.81 & 57 & R\\
    B & 0.99 & 0.37       & 1.85    & 0.87  & 2.19 & 132 & R\\
    C & 0.95 & 0.53       & 1.28    & 0.80  & 0.70 & 129 & R\\
    D & 0.76 & 0.14       & 1.93    & 0.46  & 2.12 & 165 & P\\
    E & 0.80 & 0.12       & 1.57    & 0.42  & 1.63 & 44 & P\\
    F & 0.46 & 0.13       & 1.20    & 0.37  & 2.85 & 78 & P\\
    G$^\dagger$ & 0.47 & 4.8 & 6.39   & 0.39  & - & - & P \\
\bottomrule
  \end{tabular}}

\footnotesize $^*$At the start of the simulation, subhalo A is very close to the disc
and moving away from it. Based on its velocity we estimate that it hits the disc
at $z = 1.02$.

$^\dagger$As shown in Figure~\ref{fig:traject}, subhalo G moves through the disc
plane about $10\,R_\mathrm{S}$ away from the disc centre, therefore it is not considered
as a close impact. However, thereafter subhalo G flies over the disc, largely
parallel to it, at a distance of less than $3\,R_\mathrm{S}$, which has a substantial
influence on the disc. We therefore include subhalo G in this table but omit
its impact radius and the angle of incidence. The impact redshift
  $z_\mathrm{i}$ is defined as the redshift when subhalo G is closest to the
  centre of the main halo. 
\end{table}

The column density of dark matter at $z = 1$ is shown in the top row of
Figure~\ref{fig:traject}. Outside of the main halo there is a large number of
subhaloes that may fall in at a lower redshift. A zoom-in view of the column
density of all subhaloes in the Phase-3 simulation at $z = 1$ is shown in the
middle row of Figure~\ref{fig:traject}. From all these subhaloes we select the
ones that interact with the disc based on the following criteria: {\it a)}
their impact radius $r$ is less that $5\,R_\mathrm{S}$, {\it b)} their mass exceeds $
10^9\,\mathrm{M_\odot}$. There are more subhaloes that either hit the disc plane
further out or have a lower mass, but we filter them out as they play a minor
role in the development of spiral structures in the disc as we have explicitly
checked (see e.g. Appendix~\ref{sec:abc}). The properties of 
impacting subhaloes based on our criteria above are listed in
Table~\ref{tab:1}. They hit the disc at various redshifts from $z = 1$ to $z
\sim 0.4$ and have different angle of incidence and rotation
direction. Generally, as expected, subhaloes that are further away from the
centre of the halo hit the disc at a lower redshift. We indicate their initial
locations at $z=1$ in the first two rows of Figure~\ref{fig:traject} with
different letters, and we plot their trajectories once close to the disc in the
bottom row of Figure~\ref{fig:traject}, along with the column density of the
disc. 

The impact radius $R_\mathrm{i}$ is typically between
  $0.7R_\mathrm{S}$ and $2.9R_\mathrm{S}$. The half-mass radius of the subhaloes
  at the time of maximum mass before the impact
is lower than the impact radius for all subhaloes except for subhalo C and G,
for which $R_\mathrm{hm}$ and $R_\mathrm{i}$ are comparable. However, note that at the
time of impact, the half-mass radius decreases to $10$-$40\%$ of its
  maximum value before  the impact as a result of
tidal stripping, which means that
at the time of impact, the subhaloes should be considered as objects of
relatively small size. It is therefore a reasonable estimate to consider an impacting
subhalo in our simulation  as a point  mass.

A closer look at each subhalo reveals that there are
in fact only 5 subhaloes in the 7 subhalo events, as subhalo A and C are
the same subhalo hitting the disc twice, so are subhaloes B and E. It is also
worth noting that subhalo A does not hit the disc
during the Phase-3 simulation. It goes through the disc at
$z=1.02$, before the beginning of the Phase-3 simulation, and at the start of
Phase-3, it is located very close to the disc, moving away from it. Subhalo
G does not hit the disc within $5\,R_{\rm S}$, but crosses the disc plane
further away from the disc centre. However, it is very close to the disc at
$z\sim 0.4$ passing over the disc at a distance of no more than $3\,R_{\rm
  S}$, as shown by the black curve in the bottom row of
Figure~\ref{fig:traject}.

\begin{figure*}
  \centering
  \includegraphics[width=\linewidth]{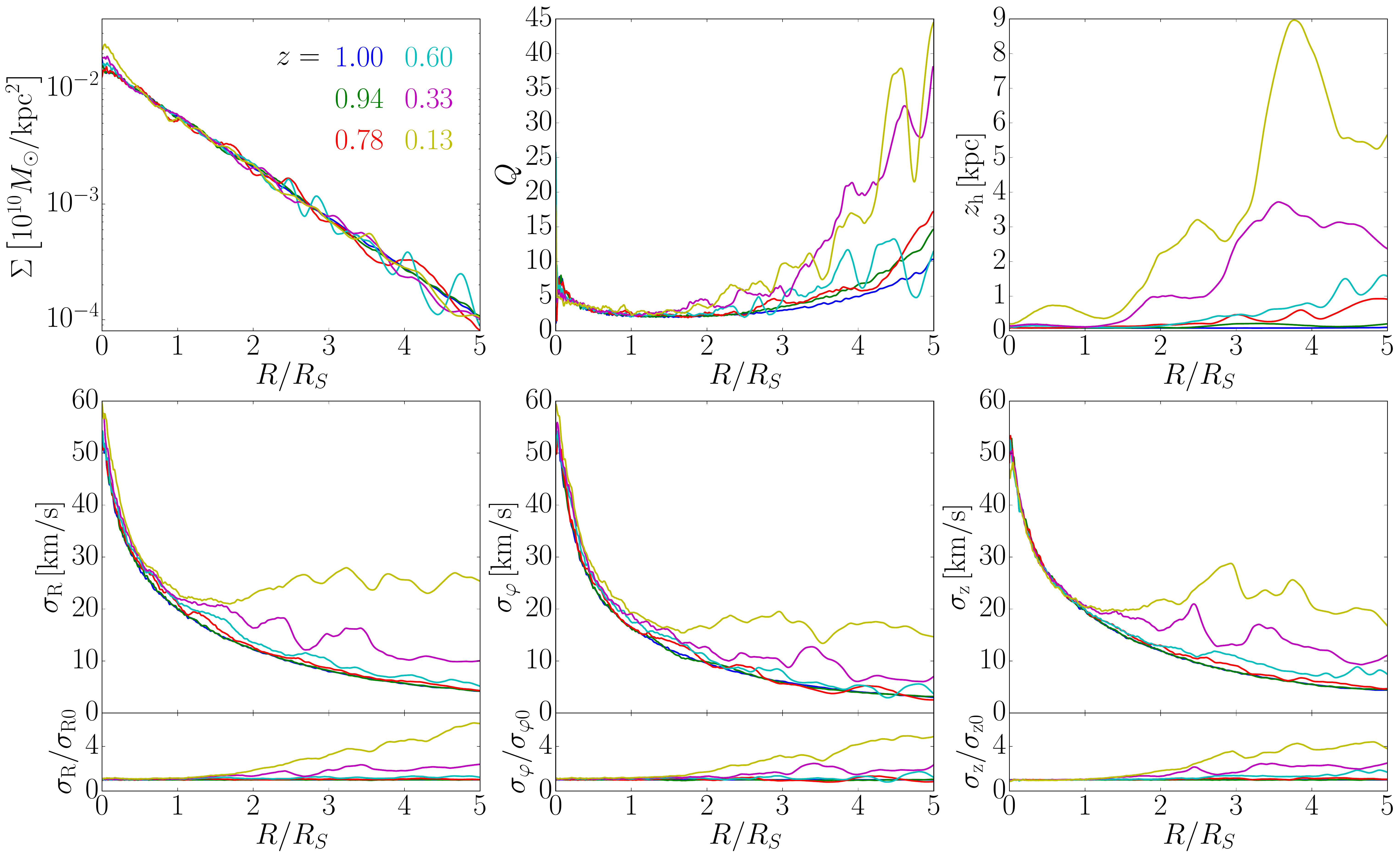}
  \caption{The time evolution of disc's surface density, $\Sigma$, Toomre's $Q$
    parameter, disc thickness, $z_\mathrm{h} = \sqrt{\left\langle z^2\right\rangle
        -\left\langle z \right\rangle^2}$, and velocity dispersion in the
    radial, azimuthal and vertical direction,  
    $\sigma_{\rm R}$, $\sigma_{\rm \varphi}$ and $\sigma_{\rm z}$,
    respectively.  The surface density of the disc fluctuates in the outer
    region and increases somewhat in the innermost region at later times. The
    Toomre's $Q$ parameter increases significantly in the outer region,
    i.e. for $R \gtrsim 2R_{\rm S}$, which can be explained by the increase of
    $\sigma_{\rm R}$. Disc also thickens considerably in the outer region
      after $z=0.5$. Aside from thickening, we find strong warps in the
      disc after $z=0.5$.}\label{fig:35}
\end{figure*}

As mentioned in Section~\ref{sec:phase3}, an analytic dark halo potential is
employed in the Phase-3 simulation instead of a live dark matter main halo. To
study its influence on the subhaloes and understand if it is biasing our
results in any way, we compared the mass loss and trajectory of each subhalo
listed in Table~\ref{tab:1} in the Phase-3 simulation with its counterpart in
the Phase-1 simulation, where a live main halo is present (but not a live
stellar disc). We find that the mass loss of each subhalo at the moment of
impact is similar between the two simulations, ranging from $50\%$ to $80\%$
(for further details see Appendix~\ref{sec:livevsstatic}). The mass loss is
mainly induced by  tidal disruption in both simulations. We also note that
due to dynamical friction, the subhaloes in the Phase-1 simulation move slower
than those in the Phase-3 simulation, which delays the time of the impact by up
to $1\,\mathrm{Gyr}$, but subhaloes have comparable velocities at impact and spend
similar amounts of time in the vicinity of the disc. Therefore, with respect to
the properties of subhaloes, using an
analytic main halo potential instead of a live dark matter main halo should
not affect our results below.

\subsection{Disc Heating in Response to Subhaloes}
\label{sec:disc-heating}

As the subhaloes interact with the disc its properties change, as shown in
Figure~\ref{fig:35}, especially at later times when the massive, fly-by
subhalo G interacts with the disc. The spherically averaged surface density of
the disc (top left panel) fluctuates in the outer region due to strong spiral
and ring structures caused by subhaloes. In the innermost region, the density
of the disc grows mildly at later times as a bar gradually forms. Top middle
panel shows the radial profile of Toomre's $Q$ parameter which is defined as
\begin{equation}
  \label{eq:toomre}
  Q=\frac{\sigma_{\rm R}\kappa}{3.36 \,G \Sigma}\,,
\end{equation}
where $\sigma_{\rm R}$ is the velocity dispersion in the radial direction,
$\kappa = \frac{2\Omega}{R}\frac{\mathrm{d}}{\mathrm{d}R}(R^2\Omega)$ is the
epicyclic frequency (with $\Omega$ being the rotation angular velocity of
stars), $G$ is the gravitational constant and $\Sigma$ is the surface
density. Toomre's $Q$ parameter quantifies how strong the swing amplification
is in the disc. We found in \citet{Hu2016} that the disc is stable to swing
amplification of the Poisson noise when $Q>1.3$ throughout the disc, thus we
ensure this is the case for our initial setup (see blue curve at $z = 1$).

In our Phase-3 simulation Toomre's $Q$ parameter increases in the outer region
of the disc after $z \sim 0.5$, which can be explained by the significant
increase of the velocity dispersion in the radial direction, $\sigma_{\rm R}$, shown
in the bottom left panel of Figure~\ref{fig:35}. The velocity dispersion in
the other two directions, $\sigma_{\rm \varphi}$ and $\sigma_{\rm z}$, also
increases similarly (see bottom middle and right panels). This is caused by
the interaction of the disc with subhalo G, as we will discuss in detail
in Section~\ref{sec:flyby}.

We quantify the thickness of the disc, $z_{\rm h}$, by computing the standard
deviation of the $z$ coordinate of stellar particles,
i.e. $z_{\rm h}=\sqrt{\left\langle z^2 \right\rangle-\left \langle  z \right
  \rangle^2}$. We find that the fly-by subhalo G leads to strong warps in the
disc, which causes the disc thickness, $z_{\rm h}$, to increase significantly at
later times, as shown in the top right panel of Figure~\ref{fig:35}. Further
smaller mass subhaloes interact with the disc in the wake of subhalo G
which increases the disc thickness further from $z = 0.33$ to $z = 0.13$.

\citet{Moetazedian2016} studied this heating effect with re-simulations
  of Aquarius and Via Lactea haloes with Milky Way-like discs.
  Compared to their results, we find similar ring-type density fluctuations. The
  disc thickening and heating rate before $z=0.5$ is moderately higher than
  found by \citet{Moetazedian2016}, which could be due to the fact that our stellar
  disc is less massive. For $z < 0.5$, we see much higher disc thickening and
  heating due to the massive subhalo G, which, interestingly, has a similar
mass to a massive subhalo in their Aq-F2 model (both $\sim 5\times
10^{10}$), which also caused a significant jump in disc thickness
and velocity dispersion. Furthermore, similarly to \citet{Moetazedian2016},
all of the examined disc properties are affected by the subhaloes primarily
in the outer regions, while in the innermost disc region, i.e. for $R <
2\,R_{\rm S}$, these global disc properties change much less. This is not the
case for the modes triggered in the disc which we discuss in the section
below.

\subsection{Modes in the Disc Triggered by Subhaloes}
\label{sec:modes}

\begin{figure*}
  \centering
  \includegraphics[width=.95\linewidth]{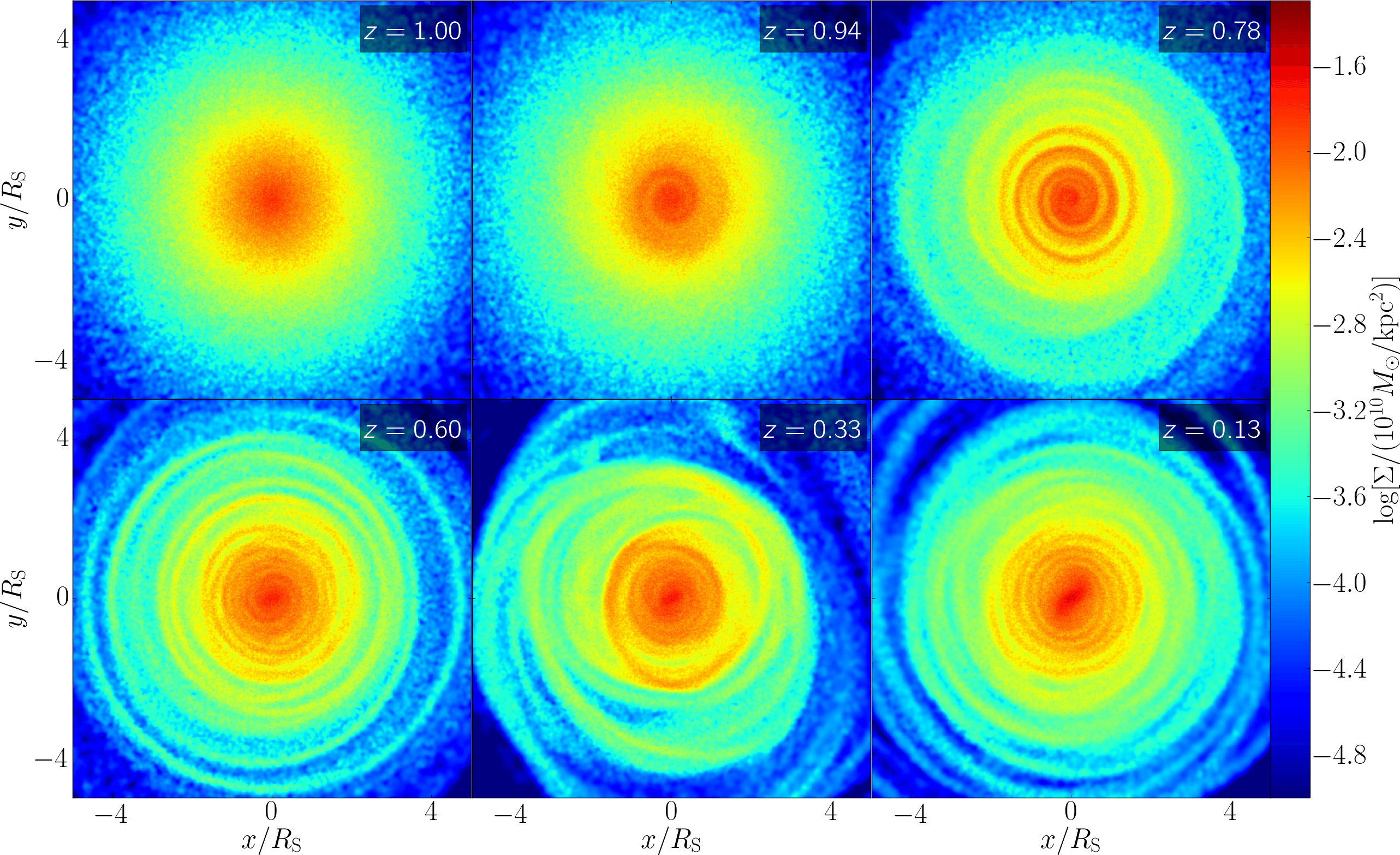}
  \includegraphics[width=.95\linewidth]{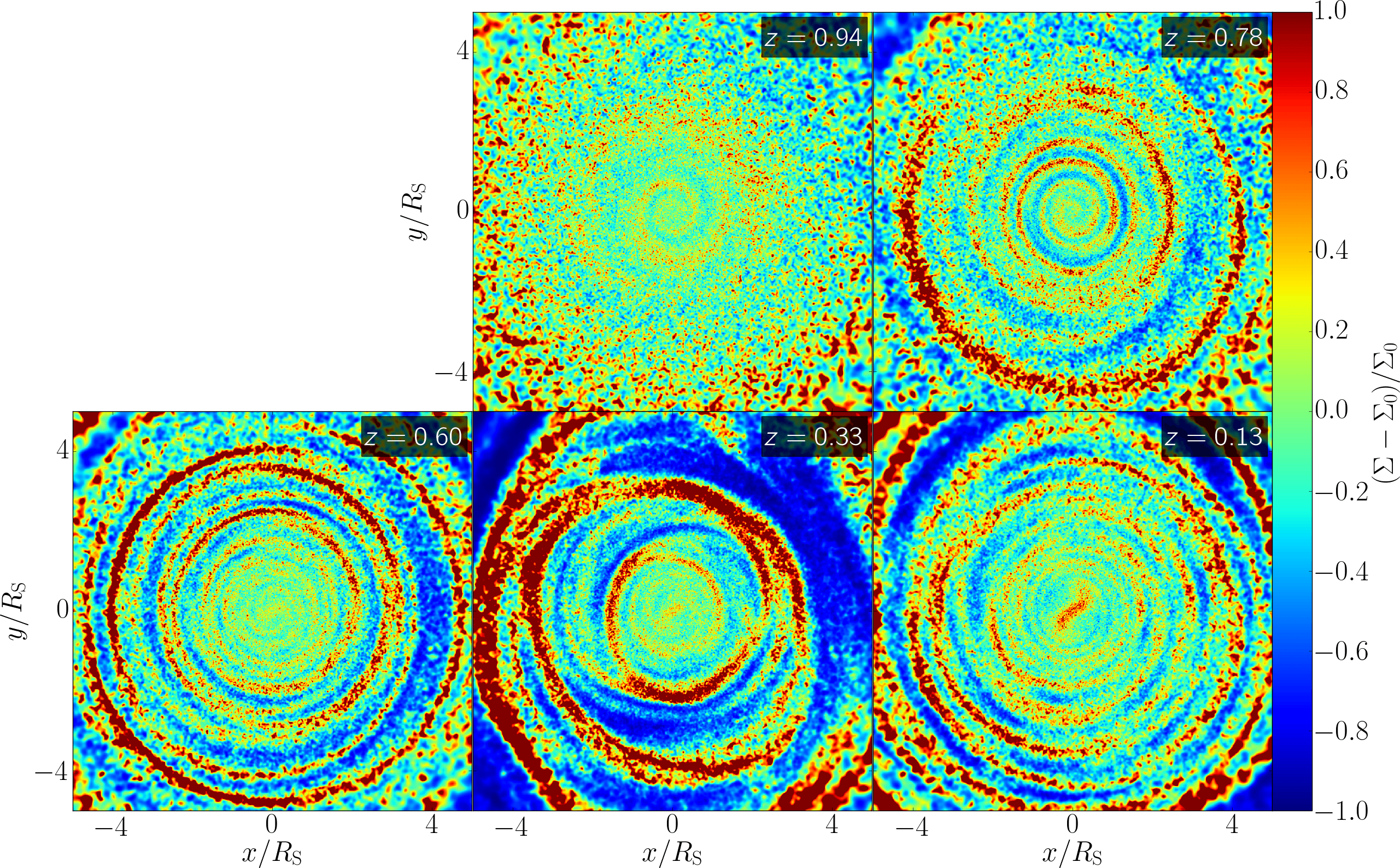}
  \caption{The evolution of the surface density $\Sigma$ (top panels) and the
    normalized residual surface density $(\Sigma-\Sigma_0)/\Sigma_0$ (bottom
    panels) at different redshifts, where $\Sigma_0$ is the surface density at
    $z = 1$. Single-armed spiral structures develop very quickly in the disc
    centre and they are counter-rotating and leading (i.e. the outer tip
      points towards the direction of the galactic rotation). These initial spiral
    structures wind up over time as the second episode of spirals
    forms in the disc at $z = 0.78$. Newly formed single-armed spiral
    structures wind up as well and become weaker at $z=0.60$, when the
    two-armed spiral structures start to become apparent in the disc
    centre. The disc stays quiet for a while until $z=0.33$, when a strong
    perturbation by subhalo G disturbs the disc in the outer
    region. This significantly warps the disc and generates the central bar. For
    $z < 0.2$ spirals wind up and weaken. The disc recovers to a quasi-stable
    state overtime and at $z = 0.13$, two-armed spiral structures are visible again
    in the central region of the disc.} 
  \label{fig:surfdens}
\end{figure*}

The interaction of subhaloes with the stellar disc leads to the development of
spirals, rings and bars. Figure~\ref{fig:surfdens} shows the time evolution
of the surface density (top panels) and the residual surface density (bottom
panels) from $z = 1$ to $z = 0.13$. By construction, the disc is free of
spiral structures at $z = 1$ and it is slightly elliptical due to the triaxial
dark matter halo. Thereafter, at least three distinct episodes of
single-armed spiral structures appear in the disc: one shortly after the
beginning of the simulation, one at $z\sim 0.78$, and one at $z\sim
0.33$. These single-armed spiral structures are counter-rotating and leading
(with respect to the direction of the galactic rotation),
which we study in detail in Section~\ref{sec:powers}. Two-armed spiral
structures also develop in the disc and are clearly visible when the
single-armed spirals wind up, namely at $z\sim 0.60$ and at $z\sim
0.13$. Both single-armed and two-armed spiral structures extend from the
centre of the disc to the edge of the disc when fully developed. While disc
remains largely axisymmetric during most of the simulated time, at $z \sim
0.33$ the interaction with the very massive subhalo G causes warps, rings,
and lopsidedness in the disc, and the formation of the central bar. At lower
redshifts, spirals wind up and weaken and the disc then evolves towards a more
axisymmetric and quasi-stable state. Note that the strength of different modes
in the disc can be quite high, especially in the outer regions, as can be seen
from the bottom panels, and we turn to quantify this next. 

\begin{figure*}
  \centering
  \includegraphics[width=.5\linewidth]{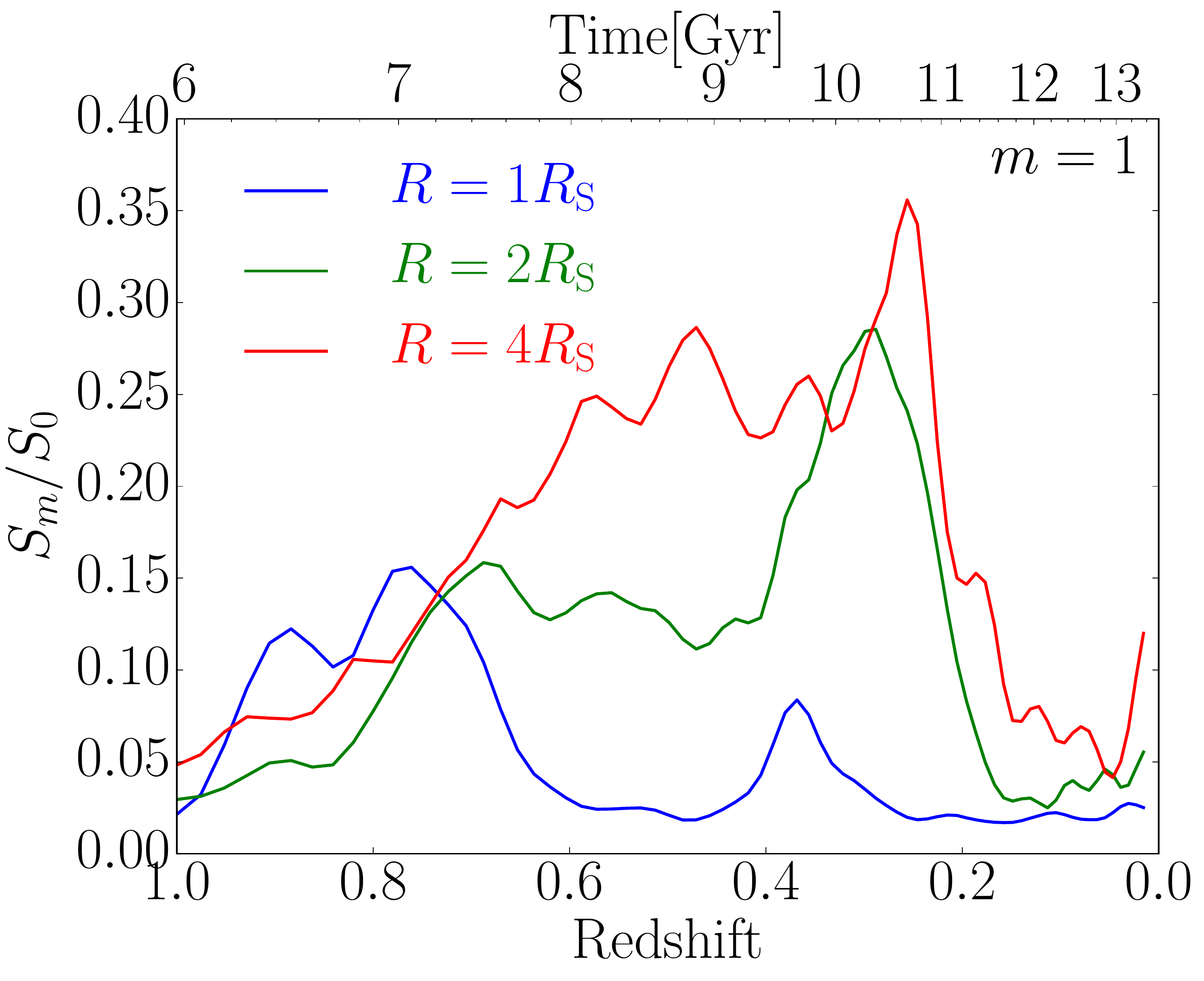}\includegraphics[width=.5\linewidth]{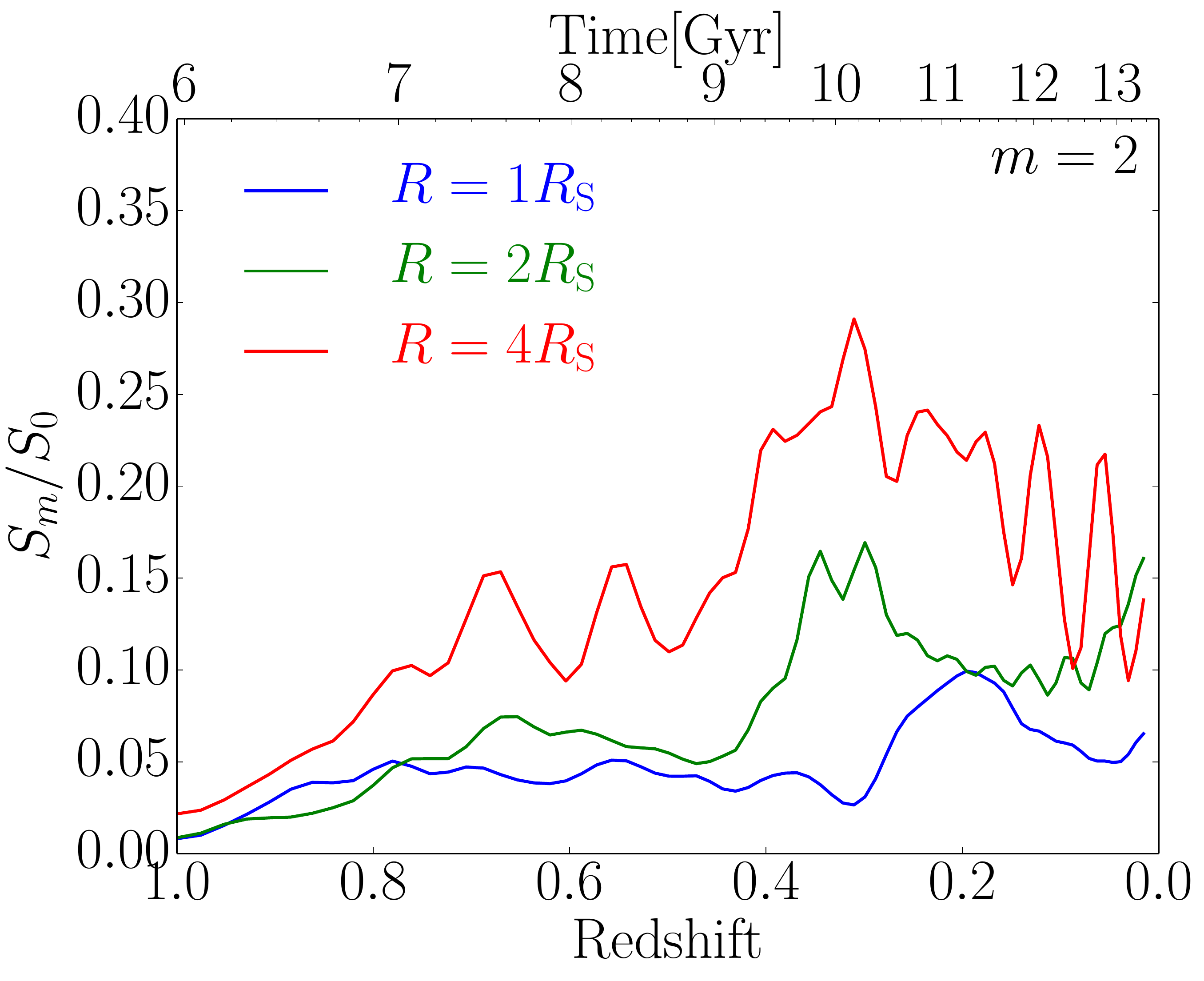}
  \caption{Strength of the modes in the disc as a function of time evaluated at $1$, $2$ and
    $4\,R_{\rm S}$ for $m = 1$ modes (left panel) and $m = 2$ modes (right
    panel). Here the relative strength $S_m/S_0$ is shown, where $S_m$ is
    defined in equation~\eqref{eq:1}. $S_0$, the strength of $m=0$ modes, is the
    spherically averaged surface
    density at radius $R$ by definition.
    For the $m=1$ case, multiple generations of modes are triggered over time,
    first one at $z\sim 1$, second one at $z\sim 0.78$, and third one at
    $z\sim 0.33$. Distinct generations of modes can be better identified in
    the inner than the outer region, as the winding time for the inner region
    is much shorter such that the different generations of modes do not
    interact significantly. The third generation of modes is also prominent
    in the $m = 2$ case.}
  \label{fig:spiral-strength}
\end{figure*}

The time evolution of the strength of $m = 1$ (left panel) and $m = 2$ (right
panel) modes is shown in Figure~\ref{fig:spiral-strength}. Here the strength of the 
modes is the relative strength of the Fourier transform of the surface density
$\Sigma(R,\theta)$ of the disc over the azimuthal coordinate $\theta$, i.e.,
\begin{equation}
\label{eq:1}
  S_{\rm m}(R)=\left|\frac{1}{2\pi}\int_\mathrm{-\pi}^{\pi}\Sigma(R,\theta)e^{-im\theta}d\theta\right|\,,
\end{equation}
where $m$ is the number of folds of the structure. For $m=0$, $S_m$ is
  simply the average surface density of a ring in the stellar disc at radius $R$. As expected, the evolution
of $m=1$ modes at $R_{\rm S}$ correlates with Figure~\ref{fig:surfdens} very well, 
showing three distinct episodes of modes from $z = 1$ to $z = 0$. The estimated life
time for each of the three episodes of spirals, measured by the full width at
  half maximum, is $\sim 1\,\mathrm{Gyr}$. Therefore,
to have persistent single-armed spirals in the stellar disc, there should be
every few Gyr massive enough subhaloes hitting the stellar disc. For the outer
region of the disc at $2R_{\rm S}$ and $4R_{\rm S}$, similar generations can also be
found, but less distinctively. This is because: {\it a)} more (smaller)
subhaloes hit the outer region of the disc, leading to a noisier evolution
history of the mode strength, and {\it b)} the life time of each episode
of modes is longer due to the much lower winding rate, as shown later
in Section~\ref{sec:powers}. $m = 2$ modes generally have lower strength and are
less distinctive, due to their even slower winding rate. Nevertheless, strong
$m = 2$ modes are generated at $z\sim 0.35$, corresponding to
the third episodes of $m = 1$ modes. We will explore the connection between
different modes generated in the disc and individual subhaloes in
Section~\ref{sec:eimp}.

\subsection{Nature of the Spiral Structures}
\label{sec:powers}

We now turn to study the nature of the spiral structures by looking at their power
spectra. As explained in detail in \citet{Hu2016}, the power spectra of the
stellar disc are the Fourier transforms of the disc's surface density over time
and the azimuthal coordinate. The power spectra of the density field are
typically plotted in the pattern speed-radius plane for each mode. If the power
contour follows the Lindblad resonance, the spiral structures are kinematic
density waves \citep{Lindblad1963}, while if the power spectra show patterns
of horizontal bars between the inner and outer Lindblad resonance, the spiral
structures are related to self-gravity \citep{sellwood2014}.

\begin{figure*}
  \centering
  \includegraphics[width=.5\linewidth]{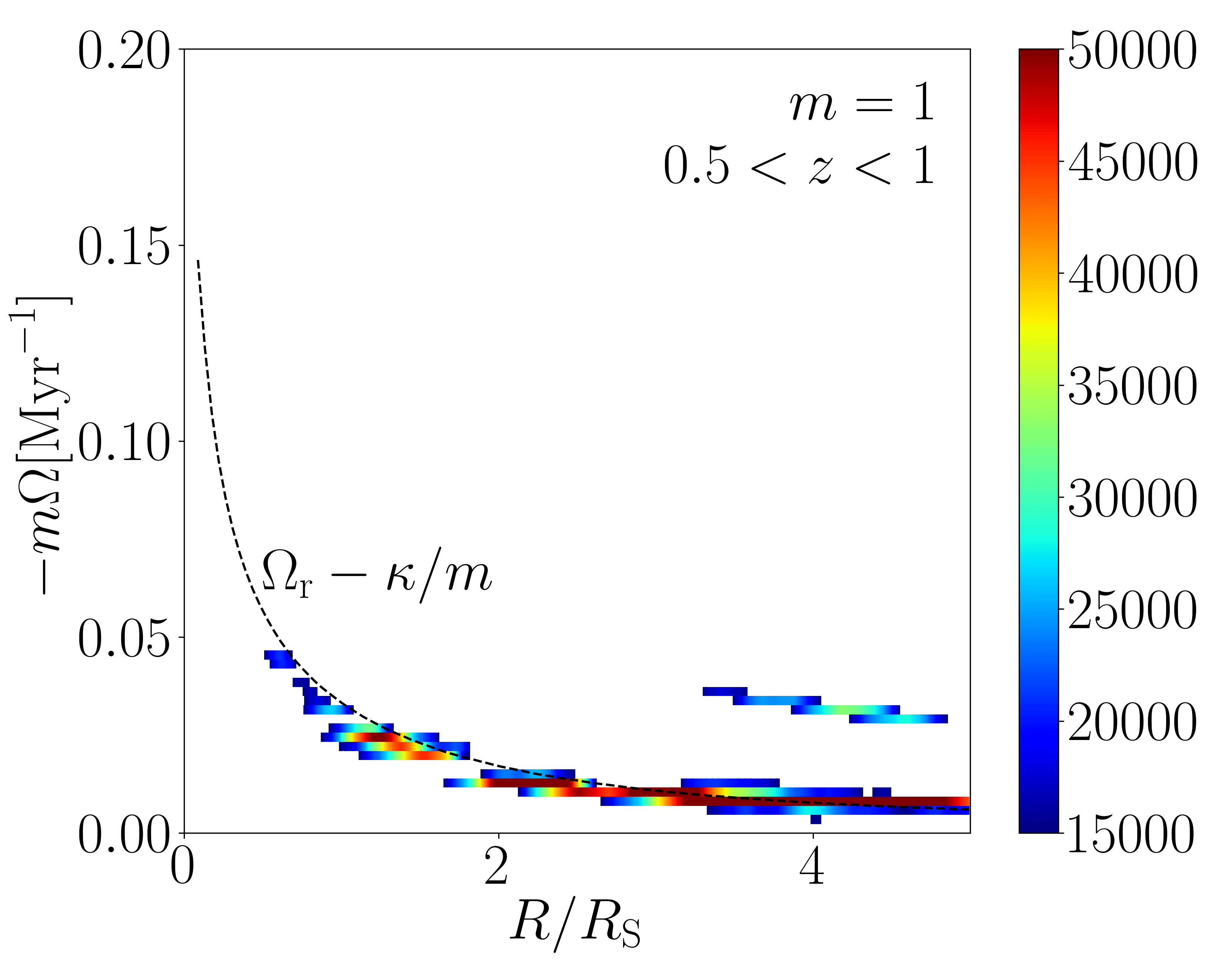}\includegraphics[width=.5\linewidth]{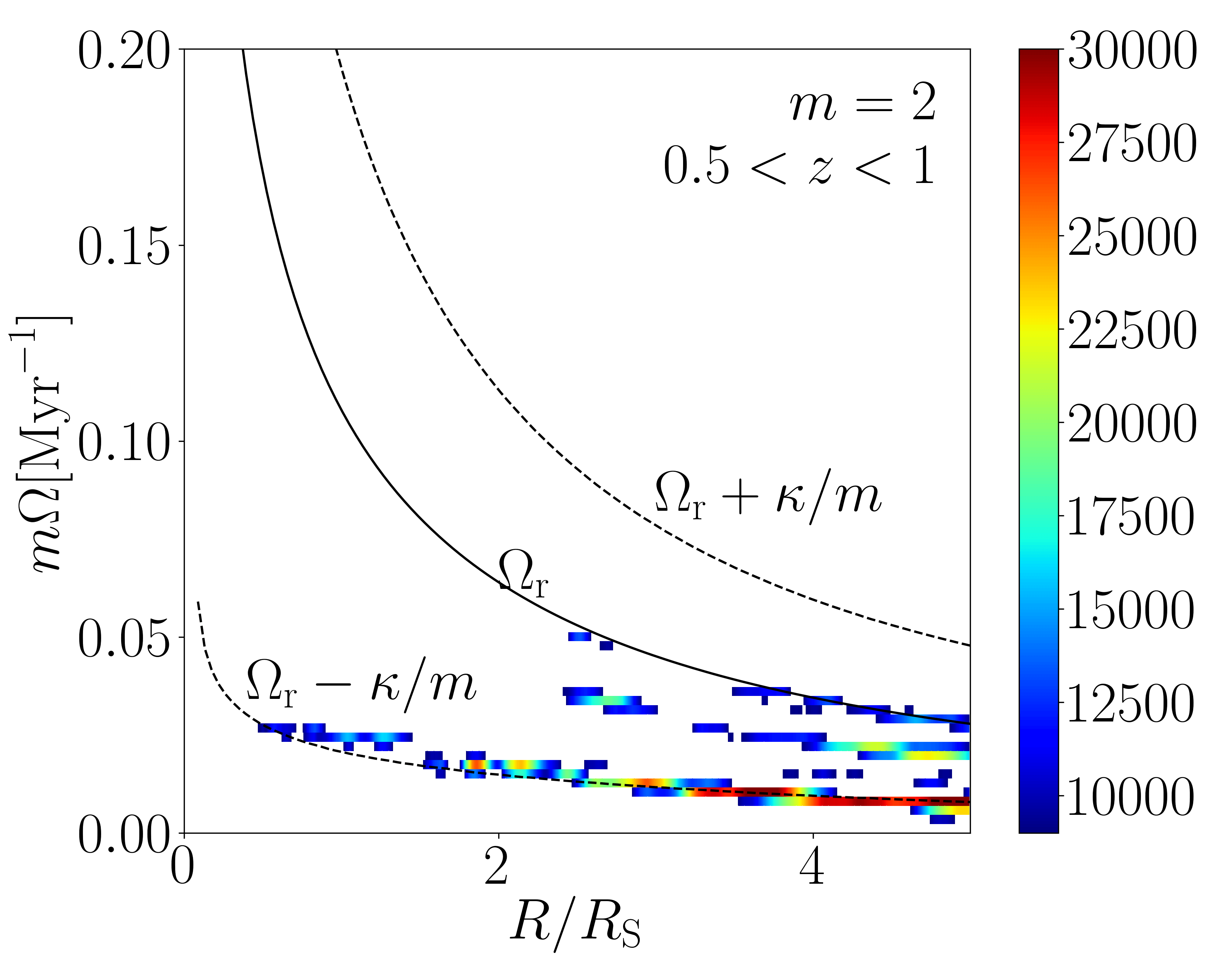}

  \includegraphics[width=.5\linewidth]{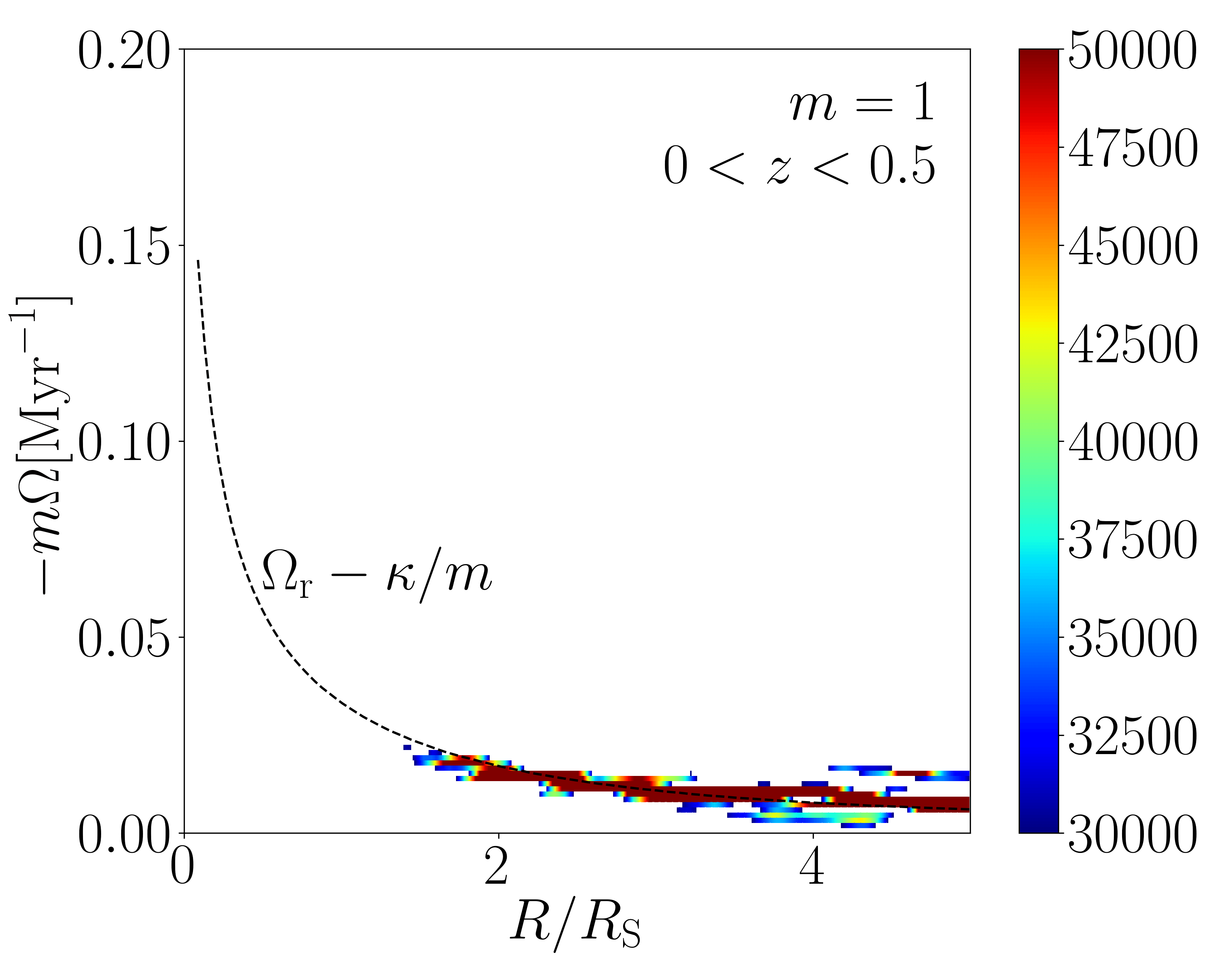}\includegraphics[width=.5\linewidth]{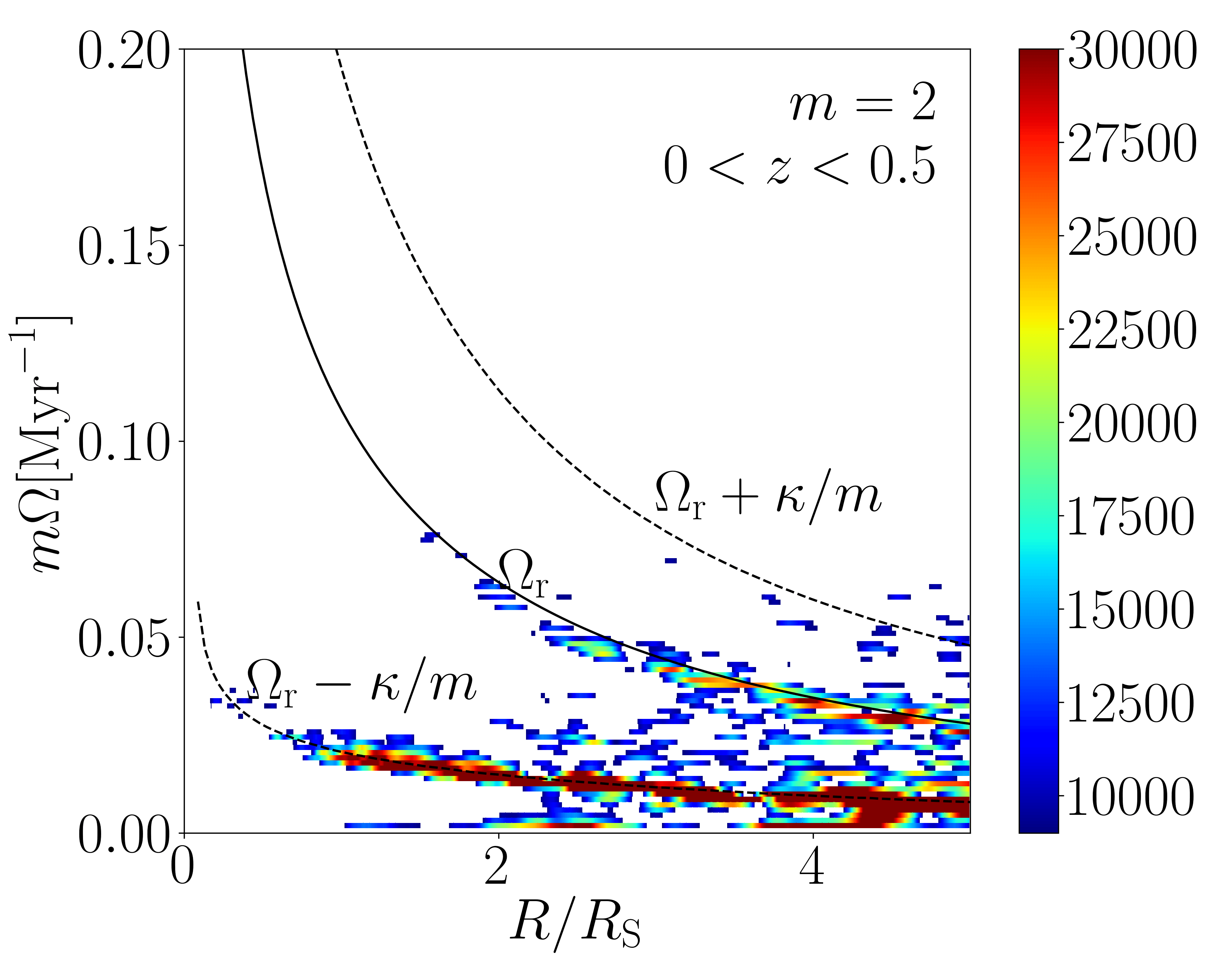}
  \caption{Power spectra of the $m=1$ (left panel) and $m=2$ (right 
    panel) modes for the time intervals from $z = 1$ to $z = 0.5$ (top panels)
    and from $z = 0.5$ to $z = 0$ (bottom panels). For $m = 1$ modes, the
    power of rotating patterns is shown in the $-m\Omega$-$R$ plane, where
    $-m\Omega$ is negative pattern speed multiplied by the number of 
    modes $m$ and $R$ is the radius from disc centre. For $m = 2$ modes, the power
    is shown in the $m\Omega$-$R$ plane. The co-rotation curve $\Omega_{\rm r}$ is
    plotted with a solid curve, while the Lindblad resonances
    $\Omega=\Omega_{\rm r}\pm\kappa/m$ are plotted with dashed curves, where
    $\Omega_{\rm r}$ is the rotating speed of stars and $\kappa$ is the epicyclic
    frequency. It can be seen that power spectra follow closely the inner
    Lindblad resonance both for $m = 1$ and $m = 2$ modes, indicating
    the kinematic density wave nature of the modes. The power of patterns
    between $z = 0.5$ and $z = 0$ is stronger than that between $z = 1$ and $z
    = 0.5$, which agrees with Figure~\ref{fig:surfdens} and
    \ref{fig:spiral-strength}.}
  \label{fig:pows}
\end{figure*}

The power spectra of our stellar disc in the Phase-3 simulation are shown in
Figure~\ref{fig:pows}. Here we plot the power spectra for $m = 1$ (left
panels) and $m = 2$ (right panels) modes for the redshift ranges of $0.5 < z <
1$ and $0 < z < 0.5$. The first redshift range corresponds to the first two
episodes of spirals, while the second redshift range corresponds to the 
third episode of spirals.
  
We plot the power for $m = 1$ modes in negative pattern speed because the
single-armed spirals are found to be counter-rotating (i.e. to have negative
pattern speed). In fact, the inner Lindblad resonance $\Omega_{\rm r} - \kappa/m$ is
also negative for $m = 1$ modes.  As shown in the left panels in
Figure~\ref{fig:pows}, at all radii the highest power follows closely  the 
inner Lindblad resonance, indicating that the modes in the disc are indeed
kinematic density waves. We should always expect grand-design single-armed
spiral structures of this kind to be counter-rotating as long as $\kappa >
\Omega_{\rm r}$. 

For $m = 2$ modes, the power spectra are weaker than in the case of $m = 1$
modes, in agreement with Figure~\ref{fig:spiral-strength}. The highest power
of $m = 2$ modes also follows the inner Lindblad resonance closely. There are 
other weaker powers away from the inner Lindblad resonance, which should be
relevant to other weaker resonances. The winding rate of spiral structures
depends on the slope of the Lindblad  resonance curve, which is much steeper
for $m = 1$ modes than the generally flat inner Lindblad curve for $m =
2$. This explains why the life time of the single-armed spiral structures,
typically less than $2\,\mathrm{Gyr}$, is lower than the life time of
two-armed spiral structures studied in \cite{Hu2016}. 

The winding rate difference also explains the inside-out fashion of
  spiral formation. As shown in Figures~\ref{fig:surfdens} and~\ref{fig:spiral-strength}, most spiral structures in
  our simulation form first in the inner region and then ``grow'' towards the outer
  regions. To illustrate this, we calculate the slope of the
  pattern speed for $m=1$, at 
  $1R_\mathrm{S}$, $\mathrm{d}\Omega_\mathrm{P}/\mathrm{d}R=0.0283
  \mathrm{Myr^{-1}}R_\mathrm{S}^{-1}$, while at $2R_\mathrm{S}$, the slope is $\mathrm{d}\Omega_\mathrm{P}/\mathrm{d}R=0.0092
  \mathrm{Myr^{-1}}R_\mathrm{S}^{-1}$. The changing rate of the tangent of the
  pitch angle $\alpha$ is 
\begin{equation}
  \label{eq:wr}
  \frac{\mathrm{d} \tan \alpha }{\mathrm{d} t}=R\frac{\mathrm{d}\Omega_\mathrm{P}}{\mathrm{d}R}\,.
\end{equation}
Therefore, the tangent of the pitch angle changes about $53\%$ faster at
$1R_\mathrm{S}$ than that at $2R_\mathrm{S}$. For comparison, it takes
$\sim 0.7\mathrm{Gyr}$ for spiral structures at $1\,R_\mathrm{S}$ to develop
with 
tripled-mass subhalo B (mentioned later in Section~\ref{sec:eimp} and Figure~\ref{fig:84}), while at
$2\,R_\mathrm{S}$, it takes $\sim 1.2\mathrm{Gyr}$, about $71\%$ longer. This explains why spiral
structures in the inner region develop faster and fade out faster.

In combination with Figure~\ref{fig:surfdens} which demonstrates that
two-armed spiral structures become prominent almost always after the winding of
single-armed spirals, we conclude that subhaloes trigger both $m = 1$
and $m = 2$ modes in the disc simultaneously, but with $m = 1$ modes initially
stronger than $m = 2$ modes. In the inner disc, the winding rate of the $m = 1$
modes is much faster than that of the $m = 2$ modes, leading to a much quicker
decrease of the strength of the $m = 1$ modes. $m = 2$ modes, winding up much
slower, become prominent after $m = 1$ modes wind up.

We also search for the self-gravitating spiral modes for higher $m$ values. The
typical strength of these modes is more than one magnitude lower than that of the
kinematic modes. This is expected as the disc has high Toomre's $Q$ parameter.

\subsection{The Impact of Each Halo}
\label{sec:eimp}
We now aim to establish a direct link between different modes generated in the
disc and the individual subhaloes that interact with the disc. We start by
studying the first generation of modes (i.e. $0.85 < z < 1$). These modes
develop immediately after the simulation starts, which is the time when
subhalo A is very close to the disc centre. As mentioned in
Section~\ref{sec:sssrs}, subhalo A is moving away from the disc at the
start of the simulation. To study the relation between subhalo A and the
first generation of modes, we restart the Phase-3 simulation with the subhalo
A removed.

To modify the mass of subhalo A and other subhaloes, we
  adopt the following procedure. We first find the most bound particle of the subhalo at
  the time of impact. The tidal stripping does not influence the
  inner core of the subhalo significantly, so this most bound particle always
  belongs to the same subhalo as we have explicitly verified.  We trace this
  particle and hence the subhalo backwards in 
  time to find the redshift where the subhalo has the
  maximum mass before the impact. This usually is the time when the subhalo is
  furthest away from the main halo. We find all particles belonging to the subhalo
  at this redshift, and change their mass in the initial conditions. We then
  restart the simulation to see the impact of changed mass.

\begin{figure}
  \centering
  \includegraphics[width=\linewidth]{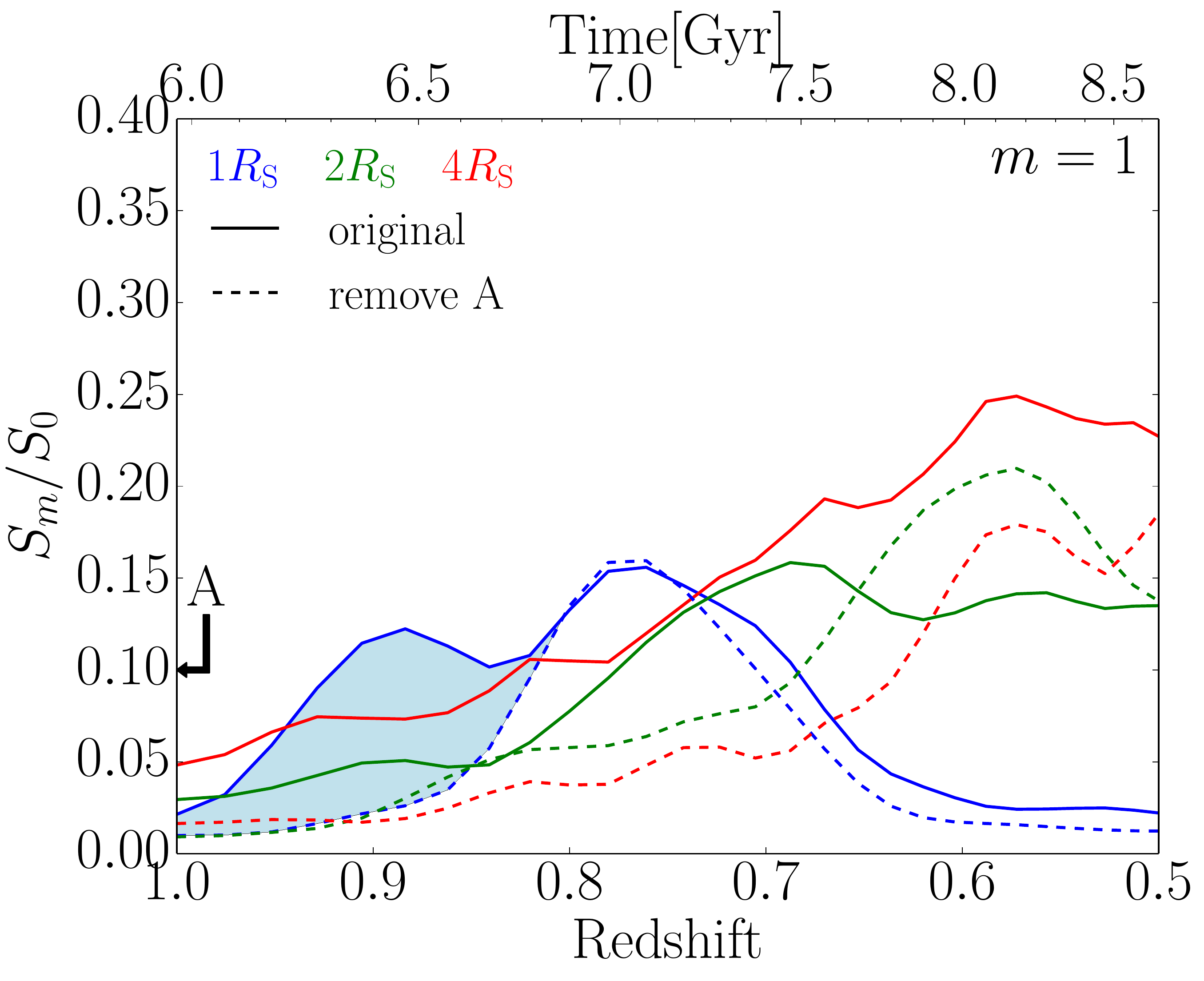}
  \caption{Comparison of the strength of the modes in the original simulation
    (continuous curves) and a simulation where subhalo A is removed
    (dashed curves). As highlighted with a light blue shade, when the subhalo
    A is removed, no prominent modes form between $z=1$ and $z=0.85$ at
    $1\,R_{\rm S}$. This is also true at $2$ and $4\, R_{\rm S}$, indicating
    that the first generation of modes is caused by the impact of subhalo A.}
  \label{fig:1-0}
\end{figure}

As highlighted by the light blue region in Figure~\ref{fig:1-0}, when the subhalo
A is removed, $m = 1$ modes before $z = 0.85$ at $1\,R_{\rm S}$ seen in the
original simulation essentially no longer develop. For $R = 2\, R_{\rm S}$ and
$R = 4\,R_{\rm S}$ we see similar decrease in mode strength, but compared to
$R = 1\,R_{\rm S}$, the influence of removing subhalo A persists longer
because the winding rate of $m = 1$ modes at $1\,R_{\rm S}$ is much higher
than that of the outer region. 

\begin{figure*}
  \centering
  \includegraphics[width=.5\linewidth]{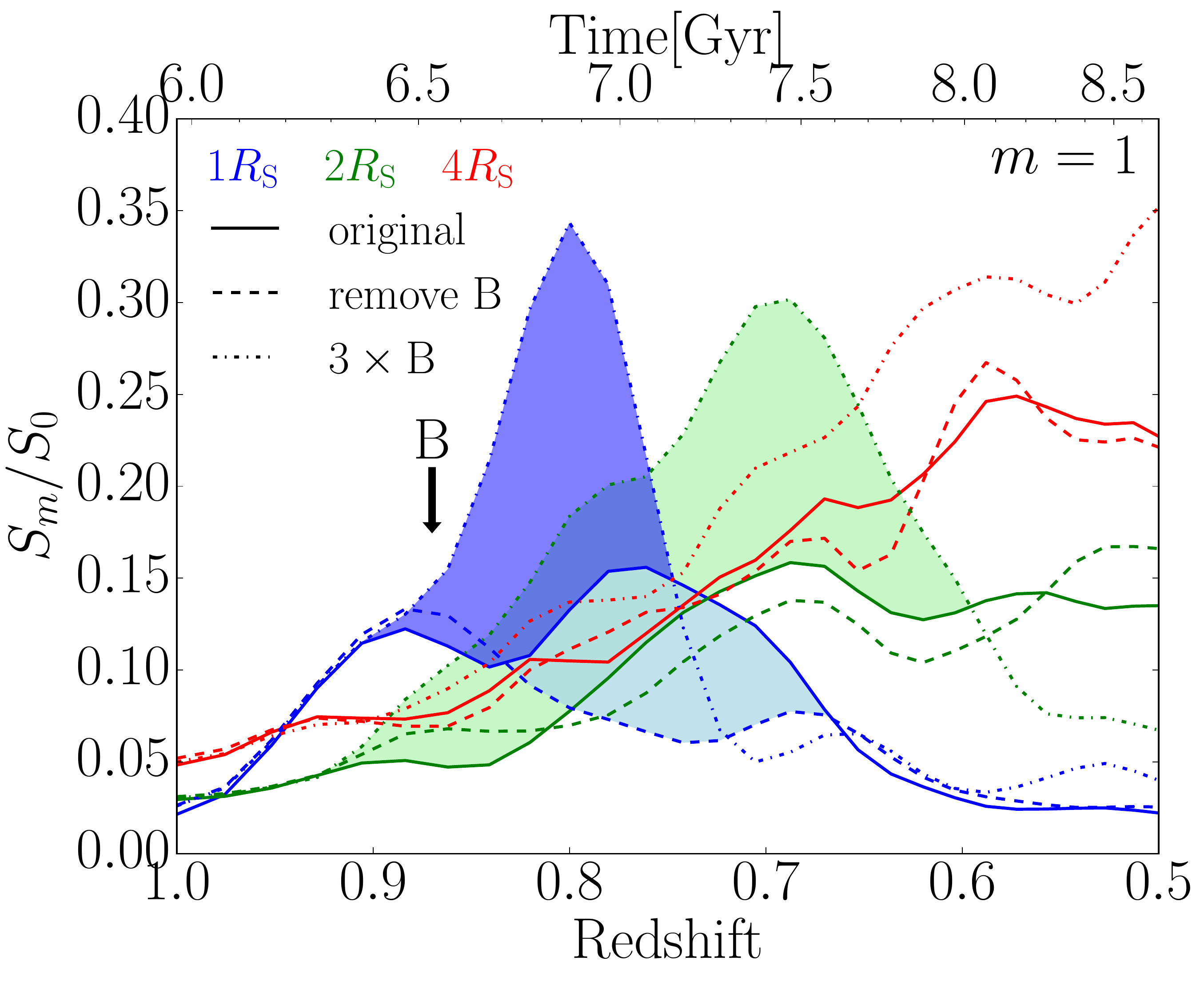}\includegraphics[width=.5\linewidth]{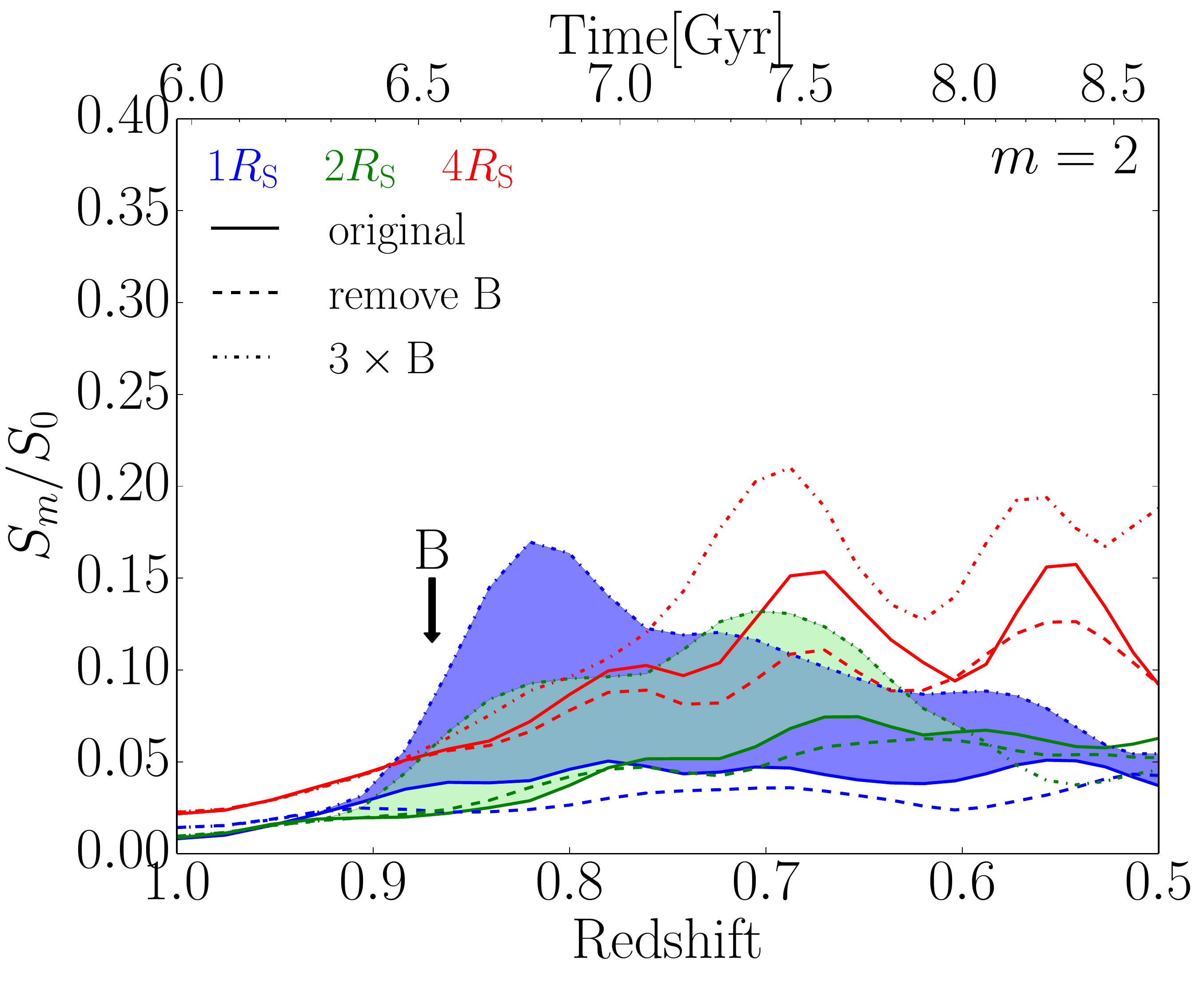}
  \caption{Comparison of the strength of the modes in the original simulation
    and simulations where either subhalo B (which hits the disc at $z=0.87$) is
     removed or its mass is tripled. When subhalo B is removed, the
    second generation of $m = 1$ modes does not form at $1\, R_{\rm S}$, as
    highlighted by the light blue region, indicating that subhalo B is the
    main cause of these modes. When the mass of subhalo B is tripled, the
    strength of both $m = 1$ and $m = 2$ modes is increased, as highlighted by
    the dark blue region for $1\,R_{\rm S}$ and by the green region for
    $2\,R_{\rm S}$. Note that in the simulation with the tripled
    mass $m = 2$ modes become dominant in central region for $z < 0.75$ (see also
    Figure~\ref{fig:3xb}).} 
  \label{fig:84}
\end{figure*}

The second generation of modes at $1\,R_{\rm S}$ develops between $z \sim 0.87$
and $z \sim 0.6$, which coincides with the time of impact of subhaloes B and
C. Recall that subhalo C is the same subhalo as A, whose influence on the
second generation of modes can be studied with Figure~\ref{fig:1-0}. By
removing subhalo A/C, the strength of the second generation of modes at $1\,R_\mathrm{S}$ reduces
only slightly for $0.6 < z < 0.75$. To study the influence of subhalo B, we 
first restart the original simulation with subhalo B removed. The result is
shown with the dashed curves in Figure~\ref{fig:84}. As highlighted by the light
blue region in the left panel, the strength of the second generation of
$m = 1$ modes at $1\,R_{\rm S}$ decreases greatly when subhalo B is
removed, while a mild decrease can be found at $2\,R_{\rm S}$. For $m = 2$
modes, even though the original spiral strength is low, we can still find a
mild decrease in the spiral strength once subhalo B is removed. Thus, the
subhalo B is the main cause of the second generation of modes with the subhalo
A/C contributing at a lower level.

\begin{figure*}
  \centering
  \includegraphics[width=\linewidth]{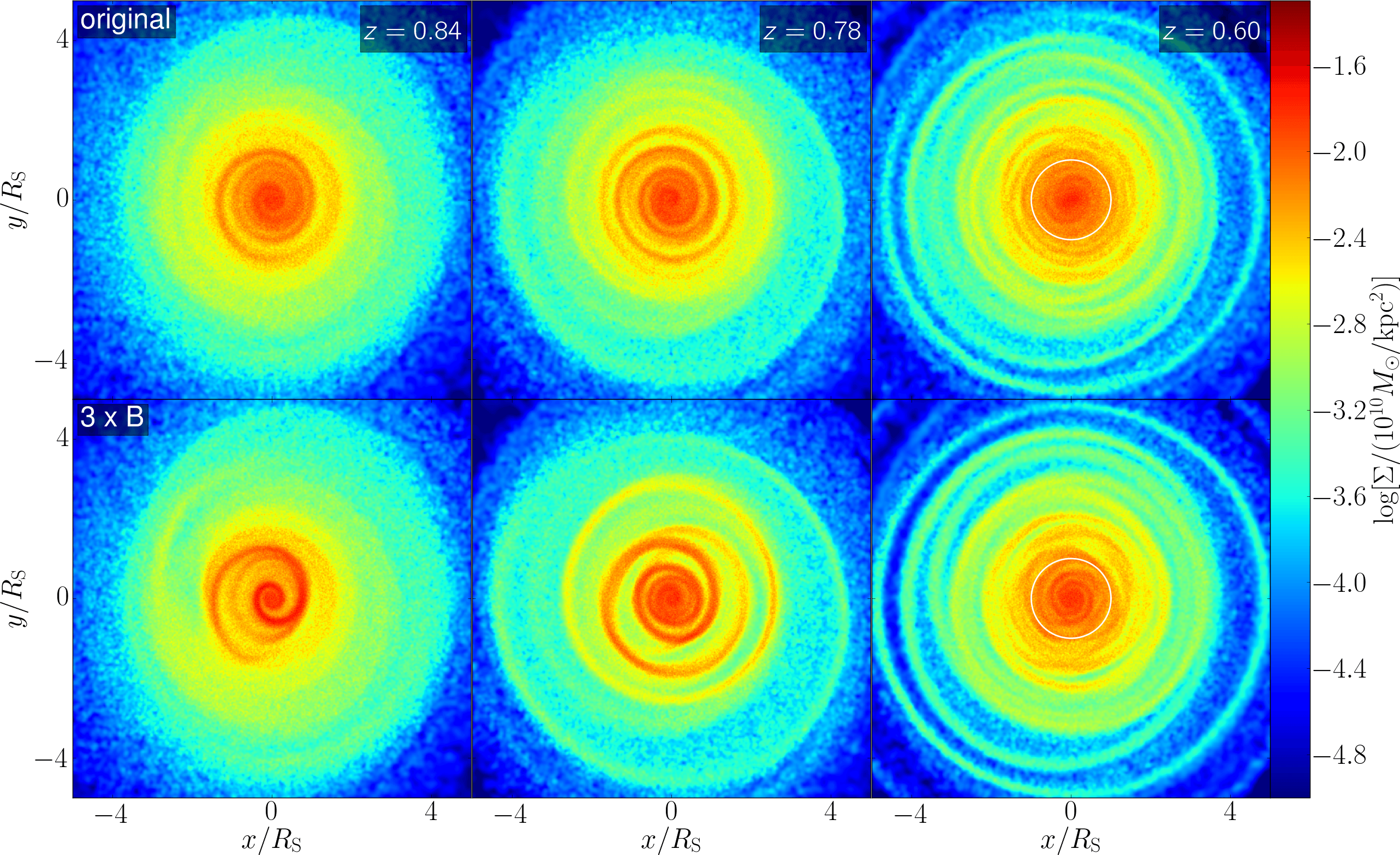}
  \caption{Surface density of the stellar disc in the original simulation (top
    panels) and the simulation where the mass of subhalo B is tripled (bottom
    panels). More massive subhalo B leads to very strong modes immediately
    after it hits the disc. Additionally to the spirals, strong ring 
    structures also develop at $z = 0.78$ in both simulations. When the
    single-armed spirals wind up at $z=0.60$, two-armed spiral structures become
  prominent in the disc centre, as highlighted by the white circles in
    the right panels. The two-armed spiral structures are more
  prominent when the mass of subhalo B is increased, in line with
  Figure~\ref{fig:84}.} 
  \label{fig:3xb}
\end{figure*}

Subhalo B starts from more than $40\,R_{\rm S}$ away from the disc centre
and is mainly responsible for the second generation of modes, offering us a
good test for studying the influence of subhalo properties, especially the
subhalo mass, on the modes triggered in the disc. We hence find all dark
matter particles that belong to subhalo B in the initial conditions,
modify their mass, and restart the simulation. We run several simulations with
different mass of subhalo B, including $1.5$, $2$ and $3$ times the
original subhalo B mass. Note that to increase the mass of subhaloes
  we increase the mass of each particle in the simulation, which increases the
  gravitational force within the subhalo, resulting in a immediate shrinkage
  in the size of the subhalo. We argue that the shrinkage is acceptable
  because \textit{a)} the maximum shrinkage in spatial scale is about
  $30\%$. At the time of impact, the half-mass radius is already small
  (typically less than $\sim 0.3\,R_\mathrm{S}$). The impact of the size
  change is therefore small; and \textit{b)} we are more interested in
  the response of the disc to subhaloes of different mass, where the spatial
  scale of a subhalo plays a minor role.

The strength of modes when the mass of subhalo B is
tripled is shown as dot-dashed curves in Figure~\ref{fig:84} (other
simulations with a lower subhalo B mass give consistent results which lie in
between the original subhalo B mass and the tripled mass). We see a clear
increase in mode strength for both $m = 1$ and $m = 2$ modes at $1\,R_{\rm S}$
(highlighted with the dark blue shade) and $2\,R_{\rm S}$ (highlighted with
the green shade). We note that for $m = 2$ modes, unlike in the 
original simulation, strong two-armed spiral structures now develop. This
confirms that although the two-armed spiral structures are very weak in the 
original simulation, they are indeed triggered by the impact of the subhalo
B. At $4\,R_{\rm S}$ the strength of modes also increases with the higher mass of
subhalo B, and the influence remains for several Gyrs.  

It is also worth noting that although both $m = 1$ and $m = 2$ modes at
different radii start to develop at almost the same time, mode strength at
$1\,R_{\rm S}$ always reaches its peak before that at $2\,R_{\rm S}$.  This is
 caused by their different winding rate. When the mass of subhalo B is
tripled, the second generation of $m = 1$ modes reaches its peak strength
$\sim 0.2\,\mathrm{Gyr}$ earlier than in the original simulation at $1\,R_{\rm
  S}$. We verify that when the mass is tripled, the trajectory of subhalo B
does not change significantly. Instead, subhalo B interact with the disc
earlier because its gravitational force field is stronger and can exert a
torque on the disc earlier. 

\begin{figure*}
  \centering
  \includegraphics[width=.5\linewidth]{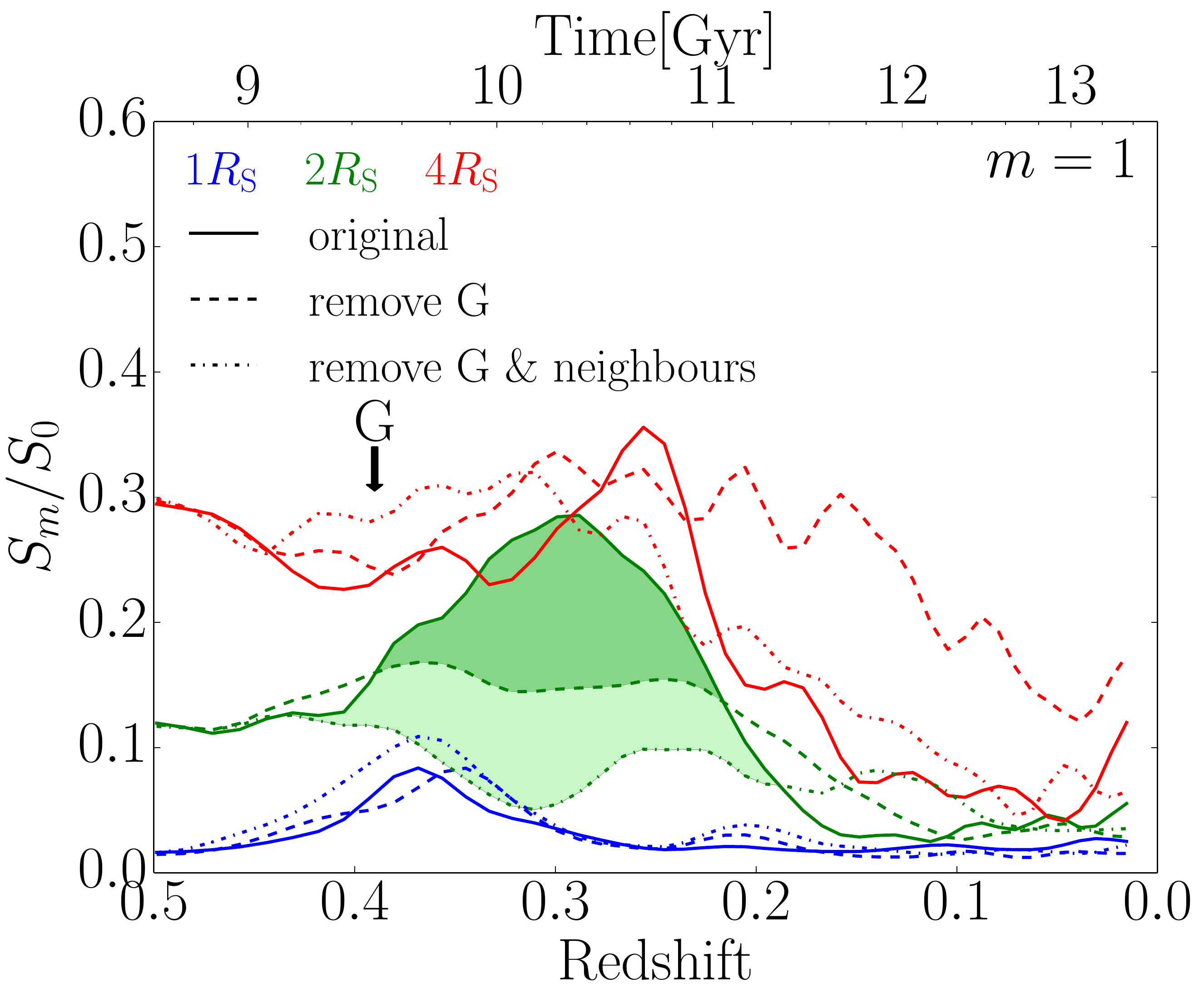}\includegraphics[width=.5\linewidth]{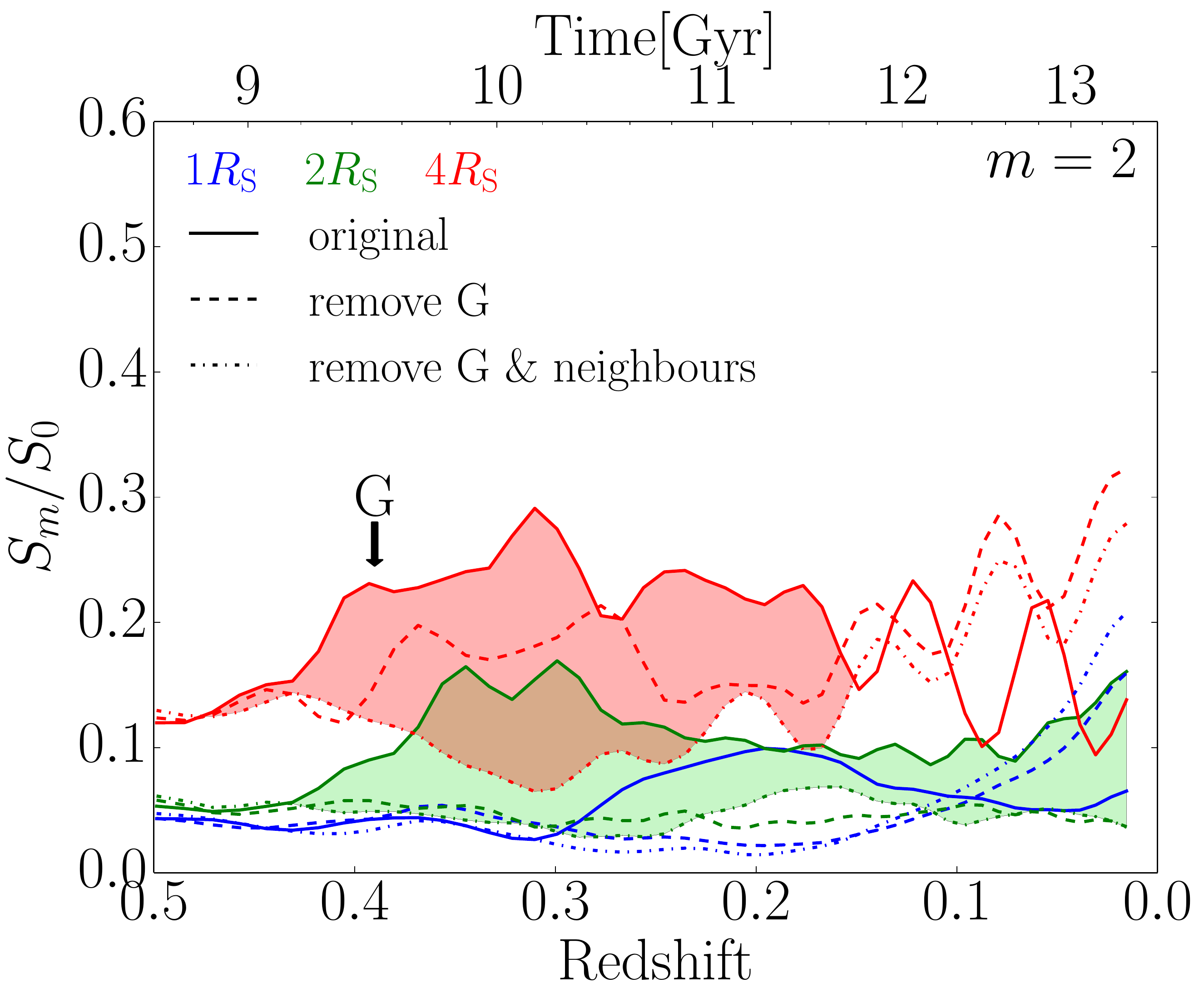}
  \caption{The strength of modes in the simulation where the fly-by subhalo G
    is removed. We either just remove subhalo G (dashed lines) or the
    surrounding subhaloes as well (dash dotted lines) which have similar
    starting positions and subsequent trajectories. The modes at $2\,R_{\rm
      S}$ are greatly reduced, as highlighted by the dark green region. When
    we remove the neighbouring subhaloes as well, the strength of the modes at
    $2\,R_{\rm S}$ and $4\,R_{\rm S}$ is reduced to a greater extend, as
    highlighted by the light green ($m = 1$ and $m = 2$) and the red ($m = 2$)
    shaded regions. The strength of modes at $1\,R_{\rm S}$ is not effected
    significantly by the removal of subhalo G, but is caused by smaller
    mass subhaloes that interact with the disc.} 
  \label{fig:fly1}
\end{figure*}

The increase in the strength of both $m = 1$ and $m = 2$ modes is also
illustrated in Figure~\ref{fig:3xb}, where a comparison of the surface density of
the original simulation and of the simulation with 3 times subhalo B mass is
shown. Similar to the original simulation, single-armed spiral structures are visible
first in the simulation with the tripled mass, while two-armed spiral
structures can be seen after the single-armed spiral structures wind up. At $z =
0.84$, the single-armed spiral structures with tripled subhalo B mass are much
more prominent than that in the original simulation, extending from the centre
of the disc to $\sim 2\,R_{\rm S}$. The single-armed spiral structures
propagate outwards from $z=0.84$ to $z=0.78$. $m = 1$ modes soon wind up, and at $z = 0.60$, very prominent
two-armed spiral structures are in the centre of the disc. 
 In the outer region, lopsided rings
also form. This effect is similar to ring galaxies. As shown by
  \citet{Lynds1976}, vertical impact of a massive point mass perturber can lead
  to such features. In fact, it has been
    shown with $N$-body simulations that energy kicks from minor mergers of
    satellites give rise to 
    radial oscillation of various frequencies in the disc, leading to ringing
    in the disc and detectable phase warping in the velocity space
    \citep{Quillen2009,Minchev2009,Gomez2012}.

We also check that subhaloes A/C and B are indeed the main cause of all modes
studied so far by removing both subhalo A and B. As shown in
Figure~\ref{fig:82}, almost all modes are removed up to $z = 0.7$. Though
there are less massive subhaloes that hit the disc during this period, the
evolution of the disc is dominated by these massive subhaloes only. We have
similarly removed subhalo D, E and F, but no prominent change in spirals is
found. We conclude the mass of these subhaloes is too small to have an
appreciable effect. This is confirmed by an additional simulation (see
Section~\ref{sec:flyby}), where we increase the mass of subhalo D by a factor
of 10 and find that it causes spiral structures in the disc. It is
  worth noting that these less massive subhaloes have a small impact on
  spirals in our simulations, although their mass is 
  larger than $10^9\,M_\odot$, the mass limit found by \cite{Pettitt2016}. This
  may be due to the fact that \cite{Pettitt2016} set up simulations with only
  one 
  subhalo at a time, while our simulations are much more dynamic with
  multiple subhaloes interacting at a time which can mask weak spirals.

\subsection{Influence of the Fly-by Subhalo}
\label{sec:flyby}

As shown in Table~\ref{tab:1} and in the bottom panels of
Figure~\ref{fig:traject}, subhalo G passes very close over the disc plane
at $z \sim 0.39$ and as we have anticipated in previous sections it perturbs
the disc significantly. We restart the original simulation with subhalo G 
removed to isolate its impact on the disc more clearly. As shown with dashed
curves in Figure~\ref{fig:fly1}, the strength of the third generation of $m =
1$ modes at $2\, R_{\rm S}$ decreases by more than $50\%$. However, unlike
direct impacting subhaloes we studied so far, no prominent difference can be
seen for $m = 1$ modes at $1\,R_{\rm S}$ at $4\,R_{\rm S}$. For $m = 2$
modes, strength at all radii decreases after removing subhalo G. When
subhalo G is closest to the disc centre, its projection on the $x-y$ plane of
the disc is $\sim 1.16\,R_{\rm S}$ away from the disc centre. As will be
discussed later, the low strength of spirals in the inner region is related to
the resonance effect. For $4\,R_{\rm S}$, the $m = 1$ mode is
not very effected because there are many more smaller subhaloes that interact
with the outer region of the disc.

We notice that some $m = 1$ and $m = 2$ modes are still present in the disc when
subhalo G is removed. We find that apart from the influence of the subhalo
G, the third generation of modes also consist of: {\it a)} the remains of
previously generated modes, {\it b)} structures generated by some smaller
subhaloes hitting the disc, and {\it c)} modes triggered by
``neighbours'' of subhalo G that also fly over the disc at a similar
time. To understand the interplay of these different effects, we remove all
smaller subhaloes that hit the disc between $z = 0.3$ and $z = 0.5$. This
removes the $m = 1$ modes at $1R_{\rm S}$. As a second experiment, we remove
subhalo G and its ``neighbours'', i.e. subhaloes whose initial positions and
trajectories are very similar to subhalo G's. The strength of the modes at
$2\,R_{\rm S}$ for $m = 1$ and $2$ and $4\,R_{\rm S}$ for $m = 2$ decreases
further, as shown in the light green and red shaded regions in
Figure~\ref{fig:fly1}. We therefore conclude that although subhalo G makes a
crucial contribution to the third generation of modes, several other subhaloes
also play a role.

\subsection{Tidally-driven Spiral Structures}
\label{sec:tidally-driven}

For all events, including direct subhalo impacts and the fly-by subhalo
interaction, single-armed spiral structures form and dissolve before
two-armed spiral structures are apparent. As explained in
Section~\ref{sec:powers}, the longevity of two-armed spirals depends on their
flat inner Lindblad resonance curve. Further to this, we find that the
interaction of each subhalo with the disc is more of an impulsive nature
rather than determined by resonances.

Previous simulation works with a single perturber, e.g.
\citet{Toomre1972,Howard1993} have found that prograde perturbers lead to $m=2$
spiral structures as the $m=2$ inner Lindblad resonance has a  prograde speed
(also see \citet{Sellwood2012} and \citet{Fouvry2015}
for the effect of resonance scattering during secular
evolution of discs.)  If subhaloes
interact with the disc for a long time, resonances with the same pattern speed
as the subhalo's rotation angular velocity will have the strongest response.
At a first glance this could be the case if we only consider subhaloes A and B. As
shown in Table~\ref{tab:1}, subhaloes A and B are both retrograde. If they
interact with the disc through resonances, modes with a negative pattern speed,
i.e. $m = 1$ modes, will be strongest, which agrees with our findings
above \citep[see also][]{Athanassoula1978,Thomasson1989}.
However, we find that the typical rotational velocity of the subhaloes
  in our simulation
  is much higher than the typical pattern speed of $m=1$ and $m=2$ modes in the
  stellar disc. As a result, the 
  interaction between  subhaloes and the disc is rather impulsive, as we show next with a 
  simulation with a massive prograde subhalo.  In this simulation, we restart the Phase-3 simulation with the
mass of subhalo D increased by a factor of $10$, where subhalo D is
prograde. If the subhalo would interact with the disc through resonances, $m =
2$ modes should be the strongest (because they have a positive pattern
speed). However, we find that single-armed spiral structures still develop first and have a higher amplitude than the $m = 2$ modes. This indicates that the
explanation of resonances does not hold in our simulations. 

Instead of resonances, we find that the strength of the spiral structures is
more related to the strength of the torque exerted by the subhalo. The
  torque strength on the stellar disc at radius $R$ exerted by a subhalo is calculated in
  the following way. For each point on the trajectory of the subhalo, we
  calculate the gravitational force field $\mathbf{F}(\mathbf{r},t)$ of the subhalo, where
  $\mathbf{r}$ is the coordinate on the disc and $t$ is time. For simplicity we
  model the subhalo as a point mass. As mentioned in Section~\ref{sec:sssrs},
  the half-mass radius of subhaloes at impact is smaller than the  impact radius, so the
point mass model should be sufficient for the estimation of torque strength. We
then calculate the torque field in normal direction of the disc $T_\mathrm{z}(\mathbf{r},t)$ as
\begin{equation}
  \label{eq:torquez}
  T_\mathrm{z}(\mathbf{r},t)= \left[ \mathbf{r}\times \mathbf{F}(\mathbf{r},t)  \right]_\mathrm{z}.
\end{equation}
To quantify
the total torque exerted by a given subhalo over its whole trajectory, we  integrate the $z$ component of the torque, $T_{\rm z}(\mathbf{r},t)$, of the
 subhalo over time to get
\begin{equation} 
  \label{eq:torquestrength}
  L(\mathbf{r})=\int_{t_{\rm 0}}^{t_{\rm f}} T_{\rm z}(\mathbf{r},t)\,\mathrm{d}t\,,
\end{equation}
where $t_{\rm 0}$ and $t_{\rm f}$ is the starting and the ending time of the impact.
As the torque quickly decreases towards 0 when the subhalo is far away, the
integral of the torque is not very sensitive to the choice of $t_{\rm 0}$ and
$t_{\rm f}$ as long as the subhalo is sufficiently far away from the disc at
both times. At a given radius $R$, the integrated torque strength $L$
  is a function of the azimuthal angle. We then calculate the $m = 1$ and the $m = 2$ component of the
torque, $L_1$ and $L_2$, with a Fourier transformation of $L$ over the
azimuthal coordinate. The spiral strength caused by each subhalo is calculated
by subtracting the corresponding peak spiral strength from the spiral strength
in the matching simulation where the subhalo is removed. 

\begin{figure}
  \centering
  \includegraphics[width=\linewidth]{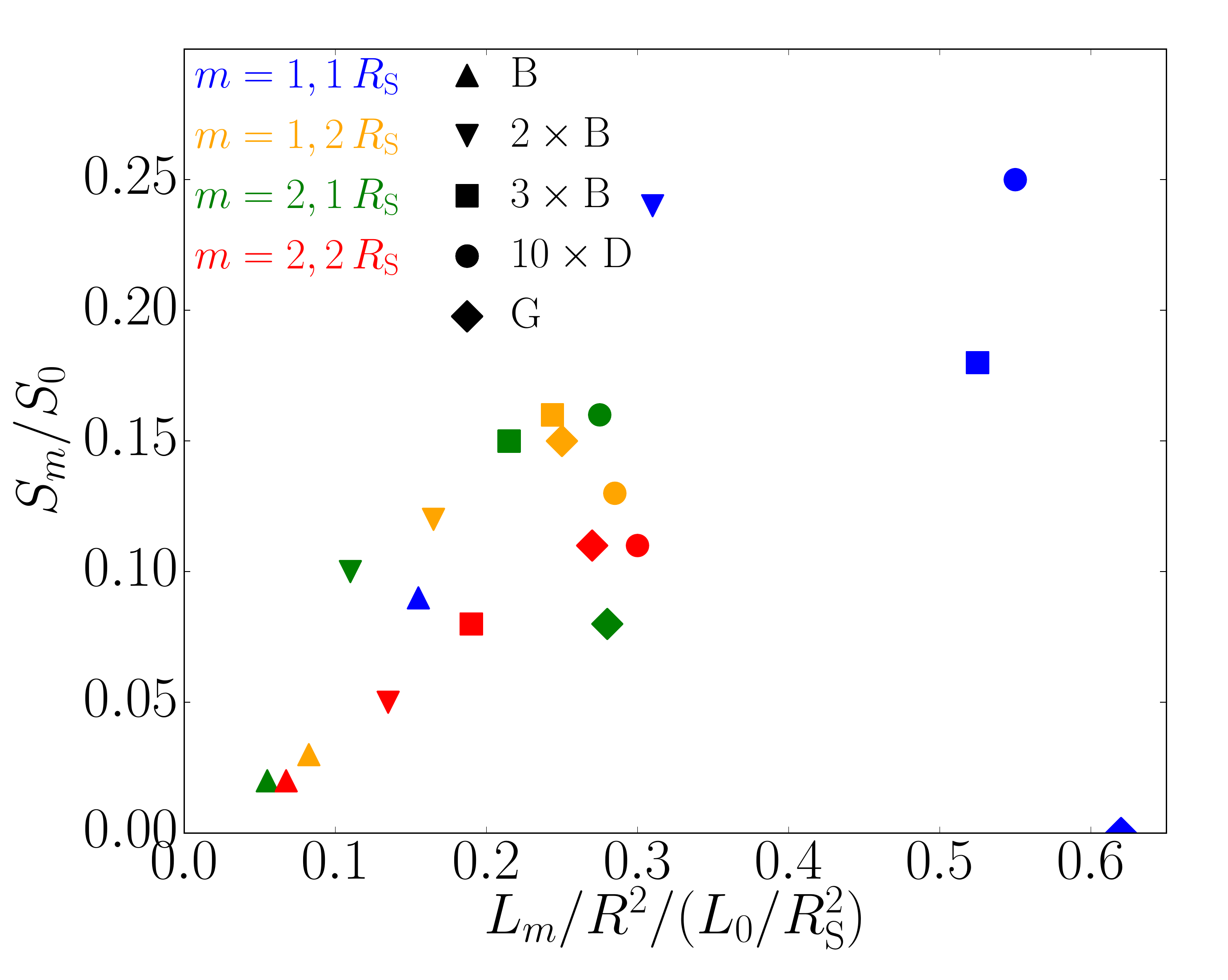}
  \caption{The relation between the relative spiral strength $S_{\rm m}$ and the
    corresponding integrated torque strength $L_{\rm m}$ for $m=1$ and $m=2$ modes,
    where $L_0 = G (10^{10} \mathrm{M_\odot}) (\mathrm{Gyr}) (\mathrm{kpc})^{-1}$ is a
    unit constant. Here we normalize the integrated torque strength $L_{\rm
      m}$ by the square of radius, $R^2$,  to account for the fact that the
    angular acceleration of particles at different radii is proportional to
    $R^{-2}$. Different $m$ modes and different radii ($1\,R_{\rm S}$ and
    $2\,R_{\rm S}$) are shown with different colours, while different symbols
    stand for different subhaloes, as listed in the legend. There is 
    a good proportional relation between the spiral strength and the
    time-integrated torque strength across different radii and different $m$ modes,
    with only a few outliers (see main text for more details).}
  \label{fig:strength-torque}
\end{figure}

We show the spiral strength as a function of the time-integrated, normalized
torque strength for several 
subhaloes, including the original subhalo B, subhalo B with double and triple
mass, subhalo D with 10 times its original mass and the fly-by subhalo G, in
Figure~\ref{fig:strength-torque}\footnote{Note that we have not included subhalo
A in this analysis as it has hit the disc before the start of the simulation,
thus making it more difficult to calculate the total torque.}. For both
$m = 1$ and $m = 2$ modes at $1\,R_{\rm S}$ and $2\,R_{\rm S}$, the spiral
strength is generally proportional to the torque strength. The only exceptions
are: {\it a)} subhalo B with tripled mass (square symbol) at $1\,R_{\rm
  S}$ for $m = 1$ modes, whose spiral strength is below the expected value due
to a saturation effect, that is, when the strength of the torque is too high,
the strength of the spiral no longer grows with the torque strength linearly
due to the constraints from the disc, and {\it b)} the fly-by subhalo G
(diamond symbol) at $1\,R_{\rm S}$ for $m = 1$ and $m = 2$ modes, where the
spiral strength is lower than the value expected for its torque strength. This
can be understood by considering the resonance explanation, as we detail
below.

The underlying reason for the strong relation between the spiral strength and
the torque strength is that the time window when the subhalo exerts a strong
torque on the disc is very short. Specifically, the time span for a subhalo to
generate torque on the disc of more than $2\%$ of the peak value is less than
$200\,\mathrm{Myr}$, except for the fly-by subhalo G, whose time span is about
$400\,\mathrm{Myr}$. The typical orbiting period of patterns in the stellar
disc ranges from $200\,\mathrm{Myr}$ (for $m = 1$ modes at $1\,R_{\rm S}$) to
$1200\,\mathrm{Myr}$ (for $m = 2$ modes at $4\,R_{\rm S}$). This means that
for most of the cases patterns evolve for no more than one orbit during the
subhalo impact.

Therefore, modes are generated and evolve in the following
way: subhaloes impact the disc impulsively, triggering different $m$ modes. Of
these, $m = 1$ and $m = 2$ modes are the strongest because the torque has a
strong $m = 1$ and $m = 2$ component. Between these two kinds of modes, $m =
1$ modes are dominant in the disc first partly because they have the
shortest winding time, while $m = 2$ modes follow and persist longer because
of their flat pattern speed curve. Another cause for stronger $m=1$ modes lies in the fact that $m=1$
  torques are generally stronger. In fact, we have calculated the strength of $m=1$ and $m=2$ components of the
torque on a ring in the stellar disc at a fixed radius when the subhalo is placed at different positions with
respect to the ring. For most regions the $m=1$ component is higher in
strength. The $m=2$ strength is higher only in a very small region close to the
ring.

 For the fly-by subhalo, the impacting time is about two times longer than
 that of the other subhaloes, comparable to or longer than the period of
 patterns at $1\,R_{\rm S}$. In this particular case, resonance effects become
 important, leading to a spiral strength that is lower than the expected value.

\section{Conclusion}
\label{sec:conclusion}
In this paper, we studied the impact of subhaloes on the stellar disc with a
series of $N$-body simulations based on the Aquarius simulations
\citep{Springel2008}. To clearly pin down the stellar disc response
to subhaloes with realistic properties extracted from the cosmological
simulations, we first performed cosmological dark matter-only simulation where
we have adiabatically introduced analytic stellar disc potential. We have then
parameterized the main dark matter halo (both in terms of its density profile
and shape and subject to the presence of the stellar disc) with an analytic
potential, fixed as a function of time, while we have represented the stellar
disc and all subhaloes found in the original Aquarius simulation with ``live''
particles. We found four massive subhaloes (subhalo A to F) that pass through
the disc, two of which (subhalo A and subhalo B) hit the disc twice (labeled
as subhaloes C and E). We also found a massive subhalo that does not impact
the disc in the innermost regions but flies over it with a very small impact
parameter (subhalo G).  

In general these subhaloes
cause disc heating, rings, warps, disc lopsidedness, a central bar, as well as
strong single- and two-armed spiral structures. There is a significant disc
heating and warping during the simulation but only at lower redshifts when
the massive fly-by subhalo G passes close to the disc. This agrees well with
other previous studies \citep[e.g.][]{Velazquez1999, Kazantzidis2009,
  Moetazedian2016}. Increase of the velocity dispersion 
leads to an increase in Toomre's $Q$ parameter, which helps stabilize
the disc from self-gravity-induced spiral structures.

Further to disc heating and warping, strong single-armed and two-armed spiral
structures develop in the disc. Generally, single-armed spiral structures are apparent
first but wind up quickly, and two-armed spiral structures become prominent after
single-armed spiral structures wind up. The winding rate of the spiral
structures can be well understood by studying the 
slope of inner Lindblad resonances, as both single-armed and two-armed
spirals turn out to be kinematic density waves whose pattern speed follows
inner Lindblad resonances. The curve of $\Omega-\kappa/m$ is steeper for $m =
1$ then for $m = 2$ at all radii, which leads to a faster winding rate of $m =
1$ modes. Nevertheless, the winding rate for the outer region of the disc is
significantly lower (both for $m = 1$ and $m = 2$ modes), such that spiral
structures can persist for several $\mathrm{Gyrs}$.

In the inner region of the stellar disc, three distinctive generations can be
found for the fast-winding single-armed spiral structures, which we attribute to
subhalo A, B and G, respectively.  We demonstrated such correlation by showing
that spiral structures are not present when we remove the corresponding
subhaloes, and that the peak strength of spiral structures in most cases
correlates very well with the torque exerted by the subhalo. This shows that
the majority of interactions between the subhaloes and the disc in our
simulations are impulsive with resonances playing a minor role. The fact that
strong spiral structures form in response to each massive subhalo, i.e. with a
mass comparable to that of the disc, may provide a way to probe the properties
of subhaloes interacting with the central galaxy. However, we caution
  that a further study taking into account the effect of main dark matter halo
  which is live (rather than static as assumed here) and considering
  stellar discs with different masses will be needed to shed light 
  on the relation between subhaloes and spiral structures.

It is worth reiterating that we have not simulated the main dark matter halo
with live particles to avoid generations of spurious modes in the disc due to
the coarse graininess of the halo (i.e. due to the Poisson noise) which is
inevitable in present state-of-the-art simulations. In reality, however,
subhaloes will suffer from a stronger dynamical friction caused by the main
dark matter halo, which may influence the evolution of the subhaloes. We have
verified that this does not effect our results in any significant way, as the
mass loss of subhaloes is similar regardless of the presence of the live halo as
is the time spent in the vicinity of the disc during which subhaloes exert most
torque and resonances. 

A live dark matter halo may also have an impact on the torque strength and on the
waves in the stellar disc. As subhaloes approach the innermost regions, the
centre of the dark matter halo and the stellar disc moves slightly in accord
with the movement of subhaloes. This may result in a smaller $m=1$ torque
(for further details see Appendix~\ref{sec:fix-main}). Additionally, in a live dark matter main halo,
  subhaloes may lead to distortions in the inner halo, which in turn act upon
  subhaloes and the disc, and damp bending waves in the disc
  \citep{Sellwood1998,Nelson1995}. As far as the warp structures are concerned,
   distortions of the inner dark matter main halo due to a passing subhalo can
   amplify the torques exerted on the disc
  \citep{Weinberg1998,Vesperini2000,Gomez2016}. We expect such effects also
  exist for spiral structures, which may even lead to stronger responses when a
  live dark matter main halo is present. In this work, we focus only on the
  impact of torques generated by the subhaloes. A study including the response
  of the dark matter main halo will require very high resolution simulations  (similar to the resolution of level-1 Aquarius
  simulation), which is beyond the scope of this paper.

Our results demonstrate a clear link between the modes in the disc and the
individual passages of massive satellites with realistic masses and orbits
extracted from cosmological simulations. Over $7$~Gyrs of cosmic time only
$2-3$ encounters are needed to continuously re-generate and sustain
grand-design spiral arms in the disc. Our stellar disc has been intentionally
setup to have a high Toomre's $Q$ profile throughout, making it possible to
isolate the effects of subhaloes only. With a disc more dominated by
self-gravity we expect that grand-design spiral arms triggered by satellites
will themselves be a source of flocculent arms thanks to swing
amplification. Hence, stellar disc interaction with satellites, within the
standard hierarchical structure formation scenario, appears as a very
promising and natural way of generating a variety of spiral structures in the
discs as well as bars, warps, tilted rings and lopsidedness.

\section*{Acknowledgements}
We thank Jim Pringle, Denis Erkal, Christophe Pichon, Lia Athanassoula and the
anonymous referee for their useful comments and advice. SH
is supported by the CSC Cambridge Scholarship, jointly funded by the China
Scholarship Council and the Cambridge Overseas Trust, and by the Lundgren
Research Award, funded by the University of Cambridge and the Lundgren Fund.  DS acknowledges
support by the STFC and ERC Starting Grant 638707 ``Black holes and their host galaxies:
co-evolution across cosmic time''. This work was performed on: DiRAC Darwin
Supercomputer hosted by the University of Cambridge High Performance Computing
Service (http://www.hpc.cam.ac.uk/), provided by Dell Inc. using Strategic
Research Infrastructure Funding from the Higher Education Funding Council for
England and funding from the Science and Technology Facilities Council; DiRAC
Complexity system, operated by the University of Leicester IT Services, which
forms part of the STFC DiRAC HPC Facility (www.dirac.ac.uk). This equipment is
funded by BIS National E-Infrastructure capital grant ST/K000373/1 and STFC
DiRAC Operations grant ST/K0003259/1; COSMA Data Centric system at Durham
  University, operated by the Institute for Computational Cosmology on behalf
  of the STFC DiRAC HPC Facility. This equipment was funded by a BIS National
  E-infrastructure capital grant ST/K00042X/1, STFC capital grant
  ST/K00087X/1, DiRAC Operations grant ST/K003267/1 and Durham University.
DiRAC is part of the National E-Infrastructure.

\bibliographystyle{mn2e}

\bibliography{./bib}

\begin{thebibliography}{}

\bibitem[\protect\citeauthoryear{{Athanassoula}}{{Athanassoula}}{1978}]{Athanassoula1978}
{Athanassoula} E.,  1978, \aap, 69, 395

\bibitem[\protect\citeauthoryear{{Bose}, {Hellwing}, {Frenk}, {Jenkins},
  {Lovell}, {Helly}, {Li}, {Gonzalez-Perez} \& {Gao}}{{Bose}
  et~al.}{2017}]{Bose2017}
{Bose} S.,  {Hellwing} W.~A.,  {Frenk} C.~S.,  {Jenkins} A.,  {Lovell} M.~R.,
  {Helly} J.~C.,  {Li} B.,  {Gonzalez-Perez} V.,    {Gao} L.,  2017, \mnras,
  464, 4520

\bibitem[\protect\citeauthoryear{Bowden, Evans \& Belokurov}{Bowden
  et~al.}{2013}]{bowden2013}
Bowden A.,  Evans N.,    Belokurov V.,  2013, Monthly Notices of the Royal
  Astronomical Society, 435, 928

\bibitem[\protect\citeauthoryear{{Boylan-Kolchin}, {Ma} \&
  {Quataert}}{{Boylan-Kolchin} et~al.}{2008}]{Boylan2008}
{Boylan-Kolchin} M.,  {Ma} C.-P.,    {Quataert} E.,  2008, \mnras, 383, 93

\bibitem[\protect\citeauthoryear{{Bullock}}{{Bullock}}{2010}]{Bullock2010}
{Bullock} J.~S.,  2010, ArXiv e-prints

\bibitem[\protect\citeauthoryear{{Collins}, {Chapman}, {Irwin}, {Martin},
  {Ibata}, {Zucker}, {Blain}, {Ferguson}, {Lewis}, {McConnachie} \&
  {Pe{\~n}arrubia}}{{Collins} et~al.}{2010}]{Collins2010}
{Collins} M.~L.~M.,  {Chapman} S.~C.,  {Irwin} M.~J.,  {Martin} N.~F.,  {Ibata}
  R.~A.,  {Zucker} D.~B.,  {Blain} A.,  {Ferguson} A.~M.~N.,  {Lewis} G.~F.,
  {McConnachie} A.~W.,    {Pe{\~n}arrubia} J.,  2010, \mnras, 407, 2411

\bibitem[\protect\citeauthoryear{DeBuhr, Ma \& White}{DeBuhr
  et~al.}{2012}]{debuhr2012}
DeBuhr J.,  Ma C.-P.,    White S.~D.,  2012, Monthly Notices of the Royal
  Astronomical Society, 426, 983

\bibitem[\protect\citeauthoryear{{Dekel} \& {Silk}}{{Dekel} \&
  {Silk}}{1986}]{Dekel1986}
{Dekel} A.,  {Silk} J.,  1986, \apj, 303, 39

\bibitem[\protect\citeauthoryear{{Diemand} \& {Moore}}{{Diemand} \&
  {Moore}}{2011}]{Diemand2011}
{Diemand} J.,  {Moore} B.,  2011, Advanced Science Letters, 4, 297

\bibitem[\protect\citeauthoryear{{Dolag}, {Borgani}, {Murante} \&
  {Springel}}{{Dolag} et~al.}{2009}]{Dolag2009}
{Dolag} K.,  {Borgani} S.,  {Murante} G.,    {Springel} V.,  2009, \mnras, 399,
  497

\bibitem[\protect\citeauthoryear{{D'Onghia}, {Springel}, {Hernquist} \&
  {Keres}}{{D'Onghia} et~al.}{2010}]{DOnghia2010}
{D'Onghia} E.,  {Springel} V.,  {Hernquist} L.,    {Keres} D.,  2010, \apj,
  709, 1138

\bibitem[\protect\citeauthoryear{D'Onghia, Vogelsberger \& Hernquist}{D'Onghia
  et~al.}{2013}]{DOnghia2013}
D'Onghia E.,  Vogelsberger M.,    Hernquist L.,  2013, The Astrophysical
  Journal, 766, 34

\bibitem[\protect\citeauthoryear{Dubinski \& Chakrabarty}{Dubinski \&
  Chakrabarty}{2009}]{dubinski2009}
Dubinski J.,  Chakrabarty D.,  2009, The Astrophysical Journal, 703, 2068

\bibitem[\protect\citeauthoryear{{Dubinski}, {Gauthier}, {Widrow} \&
  {Nickerson}}{{Dubinski} et~al.}{2008}]{Dubinski2008}
{Dubinski} J.,  {Gauthier} J.-R.,  {Widrow} L.,    {Nickerson} S.,  2008, in
  {Funes} J.~G.,  {Corsini} E.~M.,  eds, Formation and Evolution of Galaxy
  Disks Vol.~396 of Astronomical Society of the Pacific Conference Series,
  {Spiral and Bar Instabilities Provoked by Dark Matter Satellites}.
p.~321

\bibitem[\protect\citeauthoryear{{Efstathiou}}{{Efstathiou}}{1992}]{Efstathiou1992}
{Efstathiou} G.,  1992, \mnras, 256, 43P

\bibitem[\protect\citeauthoryear{{Erkal} \& {Belokurov}}{{Erkal} \&
  {Belokurov}}{2015}]{Erkal2015}
{Erkal} D.,  {Belokurov} V.,  2015, \mnras, 454, 3542

\bibitem[\protect\citeauthoryear{{Fouvry}, {Pichon}, {Magorrian} \&
  {Chavanis}}{{Fouvry} et~al.}{2015}]{Fouvry2015}
{Fouvry} J.~B.,  {Pichon} C.,  {Magorrian} J.,    {Chavanis} P.~H.,  2015,
  \aap, 584, A129

\bibitem[\protect\citeauthoryear{{Gao}, {White}, {Jenkins}, {Stoehr} \&
  {Springel}}{{Gao} et~al.}{2004}]{Gao2004}
{Gao} L.,  {White} S.~D.~M.,  {Jenkins} A.,  {Stoehr} F.,    {Springel} V.,
  2004, \mnras, 355, 819

\bibitem[\protect\citeauthoryear{{G{\'o}mez}, {Minchev}, {Villalobos}, {O'Shea}
  \& {Williams}}{{G{\'o}mez} et~al.}{2012}]{Gomez2012}
{G{\'o}mez} F.~A.,  {Minchev} I.,  {Villalobos} {\'A}.,  {O'Shea} B.~W.,
  {Williams} M.~E.~K.,  2012, \mnras, 419, 2163

\bibitem[\protect\citeauthoryear{{G{\'o}mez}, {White}, {Grand}, {Marinacci},
  {Springel} \& {Pakmor}}{{G{\'o}mez} et~al.}{2017}]{Gomez2017}
{G{\'o}mez} F.~A.,  {White} S.~D.~M.,  {Grand} R.~J.~J.,  {Marinacci} F.,
  {Springel} V.,    {Pakmor} R.,  2017, \mnras, 465, 3446

\bibitem[\protect\citeauthoryear{{G{\'o}mez}, {White}, {Marinacci}, {Slater},
  {Grand}, {Springel} \& {Pakmor}}{{G{\'o}mez} et~al.}{2016}]{Gomez2016}
{G{\'o}mez} F.~A.,  {White} S.~D.~M.,  {Marinacci} F.,  {Slater} C.~T.,
  {Grand} R.~J.~J.,  {Springel} V.,    {Pakmor} R.,  2016, \mnras, 456, 2779

\bibitem[\protect\citeauthoryear{{Grand}, {Springel}, {G{\'o}mez}, {Marinacci},
  {Pakmor}, {Campbell} \& {Jenkins}}{{Grand} et~al.}{2016}]{Grand2016}
{Grand} R.~J.~J.,  {Springel} V.,  {G{\'o}mez} F.~A.,  {Marinacci} F.,
  {Pakmor} R.,  {Campbell} D.~J.~R.,    {Jenkins} A.,  2016, \mnras, 459, 199

\bibitem[\protect\citeauthoryear{{Howard}, {Keel}, {Byrd} \& {Burkey}}{{Howard}
  et~al.}{1993}]{Howard1993}
{Howard} S.,  {Keel} W.~C.,  {Byrd} G.,    {Burkey} J.,  1993, \apj, 417, 502

\bibitem[\protect\citeauthoryear{{Hu} \& {Sijacki}}{{Hu} \&
  {Sijacki}}{2016}]{Hu2016}
{Hu} S.,  {Sijacki} D.,  2016, \mnras, 461, 2789

\bibitem[\protect\citeauthoryear{{Jiang}, {Jing}, {Faltenbacher}, {Lin} \&
  {Li}}{{Jiang} et~al.}{2008}]{Jiang2008}
{Jiang} C.~Y.,  {Jing} Y.~P.,  {Faltenbacher} A.,  {Lin} W.~P.,    {Li} C.,
  2008, \apj, 675, 1095

\bibitem[\protect\citeauthoryear{{Kazantzidis}, {Zentner}, {Kravtsov},
  {Bullock} \& {Debattista}}{{Kazantzidis} et~al.}{2009}]{Kazantzidis2009}
{Kazantzidis} S.,  {Zentner} A.~R.,  {Kravtsov} A.~V.,  {Bullock} J.~S.,
  {Debattista} V.~P.,  2009, \apj, 700, 1896

\bibitem[\protect\citeauthoryear{{Klypin}, {Kravtsov}, {Valenzuela} \&
  {Prada}}{{Klypin} et~al.}{1999}]{Klypin1999}
{Klypin} A.,  {Kravtsov} A.~V.,  {Valenzuela} O.,    {Prada} F.,  1999, \apj,
  522, 82

\bibitem[\protect\citeauthoryear{{Koposov}, {Belokurov}, {Torrealba} \&
  {Evans}}{{Koposov} et~al.}{2015}]{Koposov2015}
{Koposov} S.~E.,  {Belokurov} V.,  {Torrealba} G.,    {Evans} N.~W.,  2015,
  \apj, 805, 130

\bibitem[\protect\citeauthoryear{{Larson}}{{Larson}}{1974}]{Larson1974}
{Larson} R.~B.,  1974, \mnras, 169, 229

\bibitem[\protect\citeauthoryear{{Lindblad}}{{Lindblad}}{1963}]{Lindblad1963}
{Lindblad} B.,  1963, Stockholms Observatoriums Annaler, 22, 5

\bibitem[\protect\citeauthoryear{{Lovell}, {Frenk}, {Eke}, {Jenkins}, {Gao} \&
  {Theuns}}{{Lovell} et~al.}{2014}]{Lovell2014}
{Lovell} M.~R.,  {Frenk} C.~S.,  {Eke} V.~R.,  {Jenkins} A.,  {Gao} L.,
  {Theuns} T.,  2014, \mnras, 439, 300

\bibitem[\protect\citeauthoryear{{Lynds} \& {Toomre}}{{Lynds} \&
  {Toomre}}{1976}]{Lynds1976}
{Lynds} R.,  {Toomre} A.,  1976, \apj, 209, 382

\bibitem[\protect\citeauthoryear{{Mateo}}{{Mateo}}{1998}]{Mateo1998}
{Mateo} M.~L.,  1998, \araa, 36, 435

\bibitem[\protect\citeauthoryear{{McConnachie}, {Irwin}, {Ibata}, {Dubinski},
  {Widrow} \& e.a.}{{McConnachie} et~al.}{2009}]{McConnachie2009}
{McConnachie} A.~W.,  {Irwin} M.~J.,  {Ibata} R.~A.,  {Dubinski} J.,  {Widrow}
  L.~M.,    e.a. 2009, \nat, 461, 66

\bibitem[\protect\citeauthoryear{{Minchev}, {Quillen}, {Williams}, {Freeman},
  {Nordhaus}, {Siebert} \& {Bienaym{\'e}}}{{Minchev}
  et~al.}{2009}]{Minchev2009}
{Minchev} I.,  {Quillen} A.~C.,  {Williams} M.,  {Freeman} K.~C.,  {Nordhaus}
  J.,  {Siebert} A.,    {Bienaym{\'e}} O.,  2009, \mnras, 396, L56

\bibitem[\protect\citeauthoryear{{Moetazedian} \& {Just}}{{Moetazedian} \&
  {Just}}{2016}]{Moetazedian2016}
{Moetazedian} R.,  {Just} A.,  2016, \mnras, 459, 2905

\bibitem[\protect\citeauthoryear{{Moore}, {Ghigna}, {Governato}, {Lake},
  {Quinn}, {Stadel} \& {Tozzi}}{{Moore} et~al.}{1999}]{Moore1999}
{Moore} B.,  {Ghigna} S.,  {Governato} F.,  {Lake} G.,  {Quinn} T.,  {Stadel}
  J.,    {Tozzi} P.,  1999, \apjl, 524, L19

\bibitem[\protect\citeauthoryear{{Navarro}, {Eke} \& {Frenk}}{{Navarro}
  et~al.}{1996}]{Navarro1996}
{Navarro} J.~F.,  {Eke} V.~R.,    {Frenk} C.~S.,  1996, \mnras, 283, L72

\bibitem[\protect\citeauthoryear{{Nelson} \& {Tremaine}}{{Nelson} \&
  {Tremaine}}{1995}]{Nelson1995}
{Nelson} R.~W.,  {Tremaine} S.,  1995, \mnras, 275, 897

\bibitem[\protect\citeauthoryear{{Pe{\~n}arrubia}, {Benson}, {Walker},
  {Gilmore}, {McConnachie} \& {Mayer}}{{Pe{\~n}arrubia}
  et~al.}{2010}]{Penarrubia2010}
{Pe{\~n}arrubia} J.,  {Benson} A.~J.,  {Walker} M.~G.,  {Gilmore} G.,
  {McConnachie} A.~W.,    {Mayer} L.,  2010, \mnras, 406, 1290

\bibitem[\protect\citeauthoryear{{Pettitt}, {Tasker} \& {Wadsley}}{{Pettitt}
  et~al.}{2016}]{Pettitt2016}
{Pettitt} A.~R.,  {Tasker} E.~J.,    {Wadsley} J.~W.,  2016, \mnras, 458, 3990

\bibitem[\protect\citeauthoryear{{Purcell}, {Bullock}, {Tollerud}, {Rocha} \&
  {Chakrabarti}}{{Purcell} et~al.}{2011}]{Purcell2011}
{Purcell} C.~W.,  {Bullock} J.~S.,  {Tollerud} E.~J.,  {Rocha} M.,
  {Chakrabarti} S.,  2011, \nat, 477, 301

\bibitem[\protect\citeauthoryear{{Quillen}, {Minchev}, {Bland-Hawthorn} \&
  {Haywood}}{{Quillen} et~al.}{2009}]{Quillen2009}
{Quillen} A.~C.,  {Minchev} I.,  {Bland-Hawthorn} J.,    {Haywood} M.,  2009,
  \mnras, 397, 1599

\bibitem[\protect\citeauthoryear{{Sawala}, {Frenk}, {Fattahi}, {Navarro},
  {Bower}, {Crain}, {Dalla Vecchia}, {Furlong}, {Helly}, {Jenkins}, {Oman},
  {Schaller}, {Schaye}, {Theuns}, {Trayford} \& {White}}{{Sawala}
  et~al.}{2016}]{Sawala2016}
{Sawala} T.,  {Frenk} C.~S.,  {Fattahi} A.,  {Navarro} J.~F.,  {Bower} R.~G.,
  {Crain} R.~A.,  {Dalla Vecchia} C.,  {Furlong} M.,  {Helly} J.~C.,  {Jenkins}
  A.,  {Oman} K.~A.,  {Schaller} M.,  {Schaye} J.,  {Theuns} T.,  {Trayford}
  J.,    {White} S.~D.~M.,  2016, \mnras, 457, 1931

\bibitem[\protect\citeauthoryear{{Sellwood}}{{Sellwood}}{2012}]{Sellwood2012}
{Sellwood} J.~A.,  2012, \apj, 751, 44

\bibitem[\protect\citeauthoryear{{Sellwood} \& {Carlberg}}{{Sellwood} \&
  {Carlberg}}{2014}]{sellwood2014}
{Sellwood} J.~A.,  {Carlberg} R.~G.,  2014, \apj, 785, 137

\bibitem[\protect\citeauthoryear{{Sellwood}, {Nelson} \& {Tremaine}}{{Sellwood}
  et~al.}{1998}]{Sellwood1998}
{Sellwood} J.~A.,  {Nelson} R.~W.,    {Tremaine} S.,  1998, \apj, 506, 590

\bibitem[\protect\citeauthoryear{Springel}{Springel}{2005}]{Springel2005b}
Springel V.,  2005, Monthly Notices of the Royal Astronomical Society, 364,
  1105

\bibitem[\protect\citeauthoryear{Springel, Di~Matteo \& Hernquist}{Springel
  et~al.}{2005}]{springel2005}
Springel V.,  Di~Matteo T.,    Hernquist L.,  2005, Monthly Notices of the
  Royal Astronomical Society, 361, 776

\bibitem[\protect\citeauthoryear{{Springel}, {Wang}, {Vogelsberger}, {Ludlow},
  {Jenkins}, {Helmi}, {Navarro}, {Frenk} \& {White}}{{Springel}
  et~al.}{2008}]{Springel2008}
{Springel} V.,  {Wang} J.,  {Vogelsberger} M.,  {Ludlow} A.,  {Jenkins} A.,
  {Helmi} A.,  {Navarro} J.~F.,  {Frenk} C.~S.,    {White} S.~D.~M.,  2008,
  \mnras, 391, 1685

\bibitem[\protect\citeauthoryear{{Springel}, {White}, {Tormen} \&
  {Kauffmann}}{{Springel} et~al.}{2001}]{Springel2001}
{Springel} V.,  {White} S.~D.~M.,  {Tormen} G.,    {Kauffmann} G.,  2001,
  \mnras, 328, 726

\bibitem[\protect\citeauthoryear{{Taylor} \& {Babul}}{{Taylor} \&
  {Babul}}{2004}]{Taylor2004}
{Taylor} J.~E.,  {Babul} A.,  2004, \mnras, 348, 811

\bibitem[\protect\citeauthoryear{{Thomasson}, {Donner}, {Sundelius}, {Byrd},
  {Huang} \& {Valtonen}}{{Thomasson} et~al.}{1989}]{Thomasson1989}
{Thomasson} M.,  {Donner} K.~J.,  {Sundelius} B.,  {Byrd} G.~G.,  {Huang}
  T.-Y.,    {Valtonen} M.~J.,  1989, \aap, 211, 25

\bibitem[\protect\citeauthoryear{{Tollerud}, {Bullock}, {Strigari} \&
  {Willman}}{{Tollerud} et~al.}{2008}]{Tollerud2008}
{Tollerud} E.~J.,  {Bullock} J.~S.,  {Strigari} L.~E.,    {Willman} B.,  2008,
  \apj, 688, 277

\bibitem[\protect\citeauthoryear{{Toomre} \& {Toomre}}{{Toomre} \&
  {Toomre}}{1972}]{Toomre1972}
{Toomre} A.,  {Toomre} J.,  1972, \apj, 178, 623

\bibitem[\protect\citeauthoryear{{Velazquez} \& {White}}{{Velazquez} \&
  {White}}{1999}]{Velazquez1999}
{Velazquez} H.,  {White} S.~D.~M.,  1999, \mnras, 304, 254

\bibitem[\protect\citeauthoryear{{Vera-Ciro}, {Sales}, {Helmi}, {Frenk},
  {Navarro}, {Springel}, {Vogelsberger} \& {White}}{{Vera-Ciro}
  et~al.}{2011}]{Vera-Ciro2011}
{Vera-Ciro} C.~A.,  {Sales} L.~V.,  {Helmi} A.,  {Frenk} C.~S.,  {Navarro}
  J.~F.,  {Springel} V.,  {Vogelsberger} M.,    {White} S.~D.~M.,  2011,
  \mnras, 416, 1377

\bibitem[\protect\citeauthoryear{{Vesperini} \& {Weinberg}}{{Vesperini} \&
  {Weinberg}}{2000}]{Vesperini2000}
{Vesperini} E.,  {Weinberg} M.~D.,  2000, \apj, 534, 598

\bibitem[\protect\citeauthoryear{{Weinberg}}{{Weinberg}}{1998}]{Weinberg1998}
{Weinberg} M.~D.,  1998, \mnras, 299, 499

\end{thebibliography}

\appendix

\section{Fitting Function for the Halo}
\label{sec:ff}

In this appendix we describe our fitting function to the density distribution of
the smooth part of dark matter halo in the Phase-3 simulation. For the radial
density distribution, we 
split the profile into two regions: the main halo profile $\rho_\mathrm{I}(r)$
and the outer region profile $\rho_\mathrm{O}(r)$.  We
fit the main halo region with two Einasto profiles, namely
\begin{equation}
  \label{eq:rho}
  \rho_\mathrm{I}(r)= 
  \begin{cases}
    \rho_\mathrm{1} \exp\left[ -\frac{2}{\alpha_1}\left( \left( r/R_\mathrm{1} \right)^{\alpha_1}-1 \right)  \right]& \text{if } r<R_\mathrm{1},\\
  \rho_\mathrm{1} \exp\left[ -\frac{2}{\alpha_2}\left( \left( r/R_\mathrm{1} \right)^{\alpha_2}-1 \right)  \right]& \text{otherwise }
  \end{cases}
\end{equation}
where $R_\mathrm{1}=10\,\mathrm{kpc}$ is the scale radius,
$\rho_\mathrm{1}=9\times 10^6 \,\mathrm{M_\odot} / \mathrm{kpc}^3 $ is the scale density,
$\alpha_1=0.2$ and $\alpha_2=0.07$ are two shape parameters.

The outer region is fitted with a single Einasto profile,
\begin{equation}
  \label{eq:outer}
  \rho_\mathrm{O}=\rho_\mathrm{2} \exp\left[ -\frac{2}{\alpha_3}\left( \left( r/R_\mathrm{2} \right)^{\alpha_3}-1 \right)  \right],
\end{equation}
where $R_2=1.3\,\mathrm{Mpc}$ is the scale radius, $\rho_2=1.598\times 10^2\,
\mathrm{M_\odot}/\mathrm{kpc}^3$ is the scale density and $\alpha_3=3$ is the shape
parameter. The spherical density distribution of the  whole halo is then
\begin{equation}
  \label{eq:whh}
  \rho_\mathrm{S}(r)=\rho_\mathrm{I}(r)+\rho_\mathrm{O}(r).
\end{equation}

\begin{figure*}
  \centering
  \includegraphics[width=.5\linewidth]{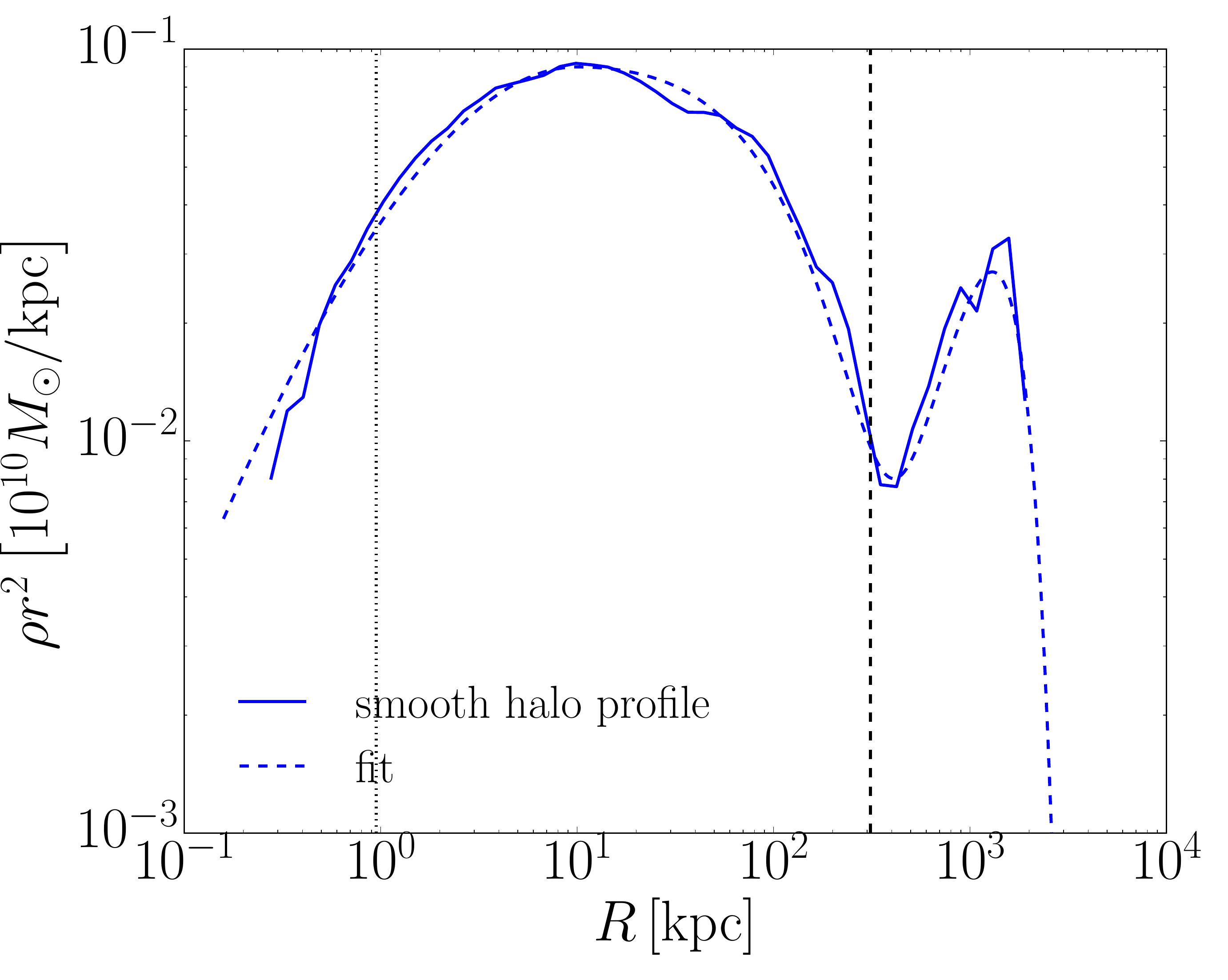}\includegraphics[width=.5\linewidth]{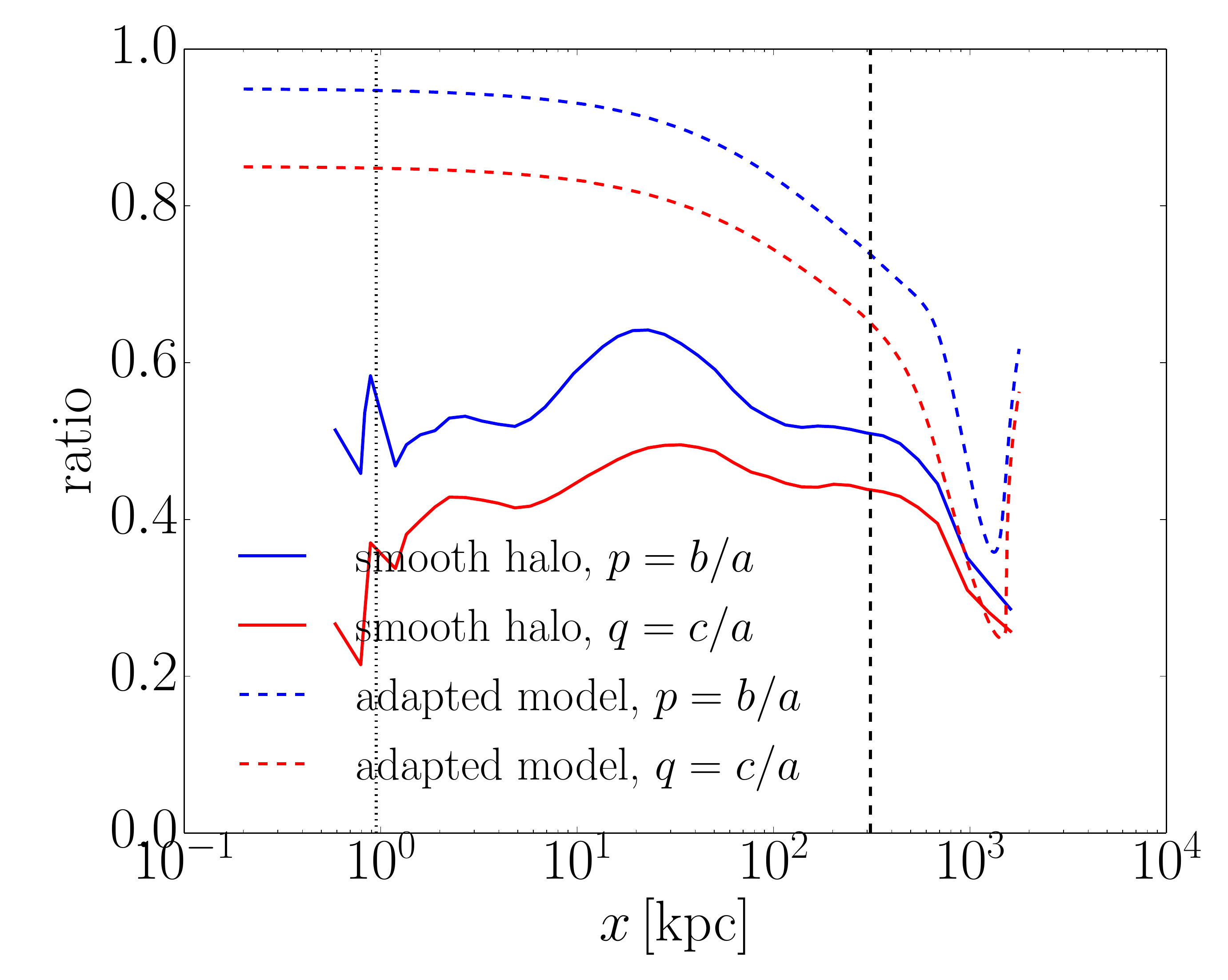}
  \caption{Fit to the smooth part of the halo in a simulation with the Aq-A
    halo and a static disc potential. The smooth part includes all particles
    except for those that become part of subhaloes from $z = 1$ to $z = 0$. In
    both plots, the solid curves represents 
    the data while the dashed curves represent our analytical model. Left: the
    $\rho r^2$ profile, where $\rho$ is the spherically averaged density. The
    valley in the density profile at $R = 400\,\mathrm{kpc}$ marks the outer
    boundary of the main halo. Right: the halo shape, i.e. the ratios of the
    three axes: $p = b/a$ and $q = c/a$, where $a$, $b$ and $c$
    are the length of three axes with $a > b > c$. We model the dark matter
    halo with triaxial models whose central region is more spherical than the
    data to account for the missing baryonic effect. The vertical dotted lines
    in both plots indicate $2.8\, \epsilon = 0.96\,\mathrm{kpc}$, where $\epsilon$
    is the softening length of dark matter particle. The vertical dashed lines
    represent the virial radius of the halo.}
  \label{fig:1}
\end{figure*}

The resulting spherically averaged halo profile, compared with the simulated
dark matter halo profile is shown in the left panel of Figure~\ref{fig:1}.

We do not set up the axis ratio profiles of our triaxial dark matter halo model by
fitting the triaxial profile of the simulated dark matter halo directly, since
we expect a much rounder inner halo profile due to missing baryonic
effects. Instead, we model the axis ratios $p = b/a$ and $q = c/a$ with the
following equation
\begin{equation}
  \label{eq:triax}
  p(r) = 0.35 + \frac{0.6}{1+\exp \left[ \log(r/r_{\rm c}) /\log(r_{\rm s})  \right]}\,,
\end{equation}
and
\begin{equation}
  \label{eq:triaxq}
  q(r) = 0.25+\frac{0.6}{1+\exp \left[ \log(r/r_{\rm c}) /\log(r_{\rm s}) \right] }\,,
\end{equation}
where $r_{\rm c} = 5.01\times 10^2\,\mathrm{kpc}$ and $r_{\rm s} = 3.16\,\mathrm{kpc}$
are the two shape parameters. The triaxial profile is implemented in the
simulation as two spherical harmonic density components as described in
Appendix~\ref{sec:ettp}. 

\section{Estimating Triaxial Density Profile}
\label{sec:ettp}

We need to relate the triaxial profile of the halo with the density profiles
representing two triaxial parts $\rho_\mathrm{T1}(r)$ and
$\rho_\mathrm{T2}(r)$ before we can use it in the simulation. The density
distribution on the three axes is related to $\rho_\mathrm{T1}$ and
$\rho_\mathrm{T2}$ through 
\begin{align}
  \label{eq:ts}
  \rho_\mathrm{x}(r)=&\rho_\mathrm{S}(r)+\frac{1}{2}\rho_\mathrm{T1}(r)+3\rho_\mathrm{T2}(r),\\
  \rho_\mathrm{y}(r)=&\rho_\mathrm{S}(r)+\frac{1}{2}\rho_\mathrm{T1}(r)-3\rho_\mathrm{T2}(r),\\
  \rho_\mathrm{z}(r)=&\rho_\mathrm{S}(r)-\rho_\mathrm{T1}(r).
\end{align}

We have to find $\rho_\mathrm{T1}(r)$ and $\rho_\mathrm{T2}(r)$ so that the
triaxial profile is satisfied, i.e.
\begin{equation}
  \label{eq:s}
  \rho_\mathrm{x}(r)=\rho_\mathrm{y}(rp(r))=\rho_\mathrm{z}(rq(r)).
\end{equation}

This equation is hard to solve analytically. We therefore approximate it
by assuming that locally the density profile follows a power law of index
$\frac{\mathrm{d}\log(\rho)}{\mathrm{d\log(r)}}$.  We
can rewrite Equation~\eqref{eq:s} as
\begin{equation}
  \label{eq:wr}
\rho_\mathrm{x}(r)=\frac{\rho_\mathrm{y}(r)}{p(r)^\alpha}=\frac{\rho_\mathrm{z}(r)}{q(r)^\alpha},
\end{equation}
where $\alpha=-\frac{\mathrm{d}\log(\rho)}{\mathrm{d\log(r)}}$ is the negative
power law index. Solving this equation yields Equation~\eqref{eq:10} and
\eqref{eq:11}. We implement this approximation and calculated the resulting
triaxial profile, as shown in the right  panel of Figure~\ref{fig:1}. The
profile is rounder in the inner region and gradually decreases to the
simulated profile at the outer region, as expected.

\section{Calculating Halo Potential from Its Density Profile}
\label{sec:cal}

The density distribution of our smooth halo follows Equation~\eqref{eq:halo}. It
is trivial to calculate the potential for the spherical component
$\rho_\mathrm{S}(r)$. To calculate potential of the triaxial part, we start from
the Poisson's equation
\begin{equation}
  \label{eq:112}
  \nabla^2 \Phi_\mathrm{T}(r,\theta,\phi)=4\pi G \rho_\mathrm{T}(r,\theta,\phi)\,.
\end{equation}
Assuming
\begin{equation}
\Phi_\mathrm{T}(r,\theta,\phi)=\sum_m\sum_l\Phi_{\mathrm{T},m}^l(r)Y_m^l(\theta,\phi)\,,
\end{equation}
and taking the density form as
\begin{equation}
\rho_\mathrm{T}(r,\theta,\phi)=\rho_\mathrm{T}(r)Y_2^{l_0}(\theta,\phi)\,,
\label{eq:www}
\end{equation}
where $l_0$ can be $0$ and $2$ for the two spherical harmonic functions used in
our model. We can easily find that $\Phi_{\mathrm{T},m}^l(r)$ is non-zero if
and only if $m=2$ and $l=l_0$. For simplicity we let
$\Phi_\mathrm{T}(r)=\Phi_{\mathrm{T},2}^{l_0}(r)$. We find 
\begin{equation}
  \label{eq:12}
  \frac{1}{r^2}\frac{\partial}{\partial r}\left( r^2\frac{\partial}{\partial r} \Phi_\mathrm{T}(r) \right)-\frac{6}{r^2}\Phi_\mathrm{T}(r)=4\pi G \rho_\mathrm{T}(r)\,.
\end{equation}
The second term on LHS is contributed by $\nabla ^2
Y_2^l(\theta,\phi)$. Green's function for a delta density function at $r_0$,
i.e. $\rho_\mathrm{T}(r)=\delta(r-r_0)$, is
\begin{equation}
  \label{eq:13}
  \mathcal{G}(r,r_0)=-\frac{1}{5}\frac{r_<^2}{r_>^3}\,,
\end{equation}
where $r_<$ and $r_>$ is the smaller and larger value in $r$ and $r_0$,
respectively. The solution to the equation is hence
\begin{equation}
  \label{eq:14}
  \Phi_\mathrm{T}(r)=4\pi G\int_0^\infty \mathcal{G}(r,r_0)\rho_\mathrm{T}(r_0)r_0^2\,\mathrm{d}r_0\,.
\end{equation}
where we convolve the density profile $\rho_\mathrm{T}(r_0)$ over the radius
$r_0$ from 0 to $\infty$ with Green's function $\mathcal{G}(r,r_0)$.
The potential and the force of the halo can then be calculated accurately with a
1-dimensional integration  in the simulation.

\section{The Artificial Torque caused by the Fixed Dark Matter Main Halo}
\label{sec:fix-main}

\begin{figure*}
  \centering
  \includegraphics[width=\linewidth]{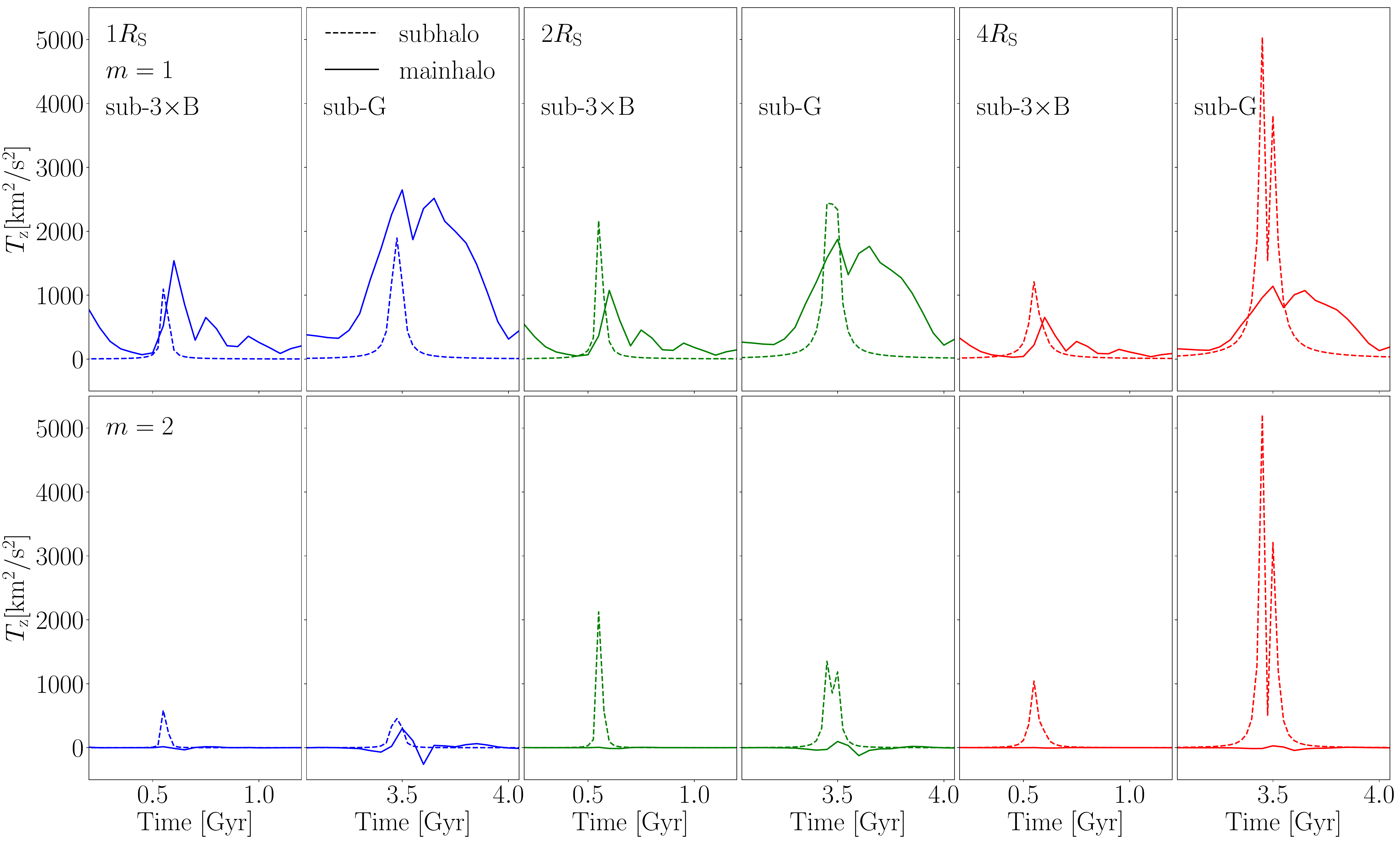}
  \caption{Comparison of torques generated by the displaced main halo and
    subhaloes $3\times$B and G. The torques at $1R_\mathrm{S}$, $2R_\mathrm{S}$
     and $4R_\mathrm{S}$ of $m=1$ (top panels) and
    $m=2$ (bottom panels) are shown. The torque generated by the main halo is
    shown with solid curves while the torque from subhaloes is shown with dashed
    curves. For $m=1$ modes at $1R_\mathrm{S}$, the artificial torque from the
    displaced 
    main halo is comparable to the torque from the subhaloes, while 
    in the outer regions it becomes smaller. The torque from the main
    halo is negligible for $m=2$ modes.}
  \label{fig:dmp}
\end{figure*}

As a subhalo approaches the innermost region of the system, the
  centre of both the stellar disc and the dark matter main halo may noticeably
  move in response to the gravitational force from the subhalo. In our
  simulation, however, the main dark matter halo is included as a fixed
  potential, and only the stellar disc responds to the approaching
  subhalo. This leads to a displacement between the main halo and the 
  disc and hence an artificial torque from the main dark
  matter halo.

In this appendix we estimate the upper limit on such a torque. We
  first calculate the displacement of the disc from its original location. We
  find that for most of the time, the disc moves no further than
  $0.3\mathrm{kpc}$ ($\sim 0.1R_\mathrm{S}$) from its original location, with
  the only exception of a displacement of $0.82\mathrm{kpc}$ ($\sim
  0.26R_\mathrm{S}$) when the subhalo G flies over the disc. Note that in
  reality only the inner region of the dark matter
  main halo should move in a similar way to that of the disc centre. Therefore
  in our simulation, the artificial torque results mostly from the fixed inner
  region of the main halo. Nevertheless, for a rough estimate of the upper
  bound on such an effect, we consider the main halo as a whole and calculate
  the torque due to its displacement with respect to the disc using
equation~\eqref{eq:torquez}.

The comparison of this torque, which is highest
in the case of the passage of subhalo G, and that caused by subhalo G itself
is shown in Figure~\ref{fig:dmp}. We also show the results for subhalo B with
three times its original mass for comparison. Due to the triaxiality of the
main halo, there is a non-zero $m=2$ torque even when the disc is not
displaced. For a better comparison, we plot the difference from that non-zero
initial torque instead of the absolute value for $m=2$ torques.

For $m=2$ modes, the torque from the main halo is generally much smaller
compared to that from subhaloes $3\times$B and G, except at $1R_\mathrm{S}$ where torques from
both are weak. Therefore our results on $m=2$ spiral structures should be in
general unaffected by the artificial torque from the fixed halo. For $m=1$
modes at $4R_\mathrm{S}$ the contribution from the displaced halo is weak as
well, but it grows in significance moving inwards, and at $1R_\mathrm{S}$ it is
comparable to the 
torque caused by subhaloes $3\times$B and G. Note however as mentioned before,
the torque 
calculated here is likely the upper bound both in terms of the peak value and
the non-zero torque width.

We further find that the phase of the torque from
the displaced main halo follows closely that from subhaloes, therefore
enhancing the $m=1$ spiral structures in our simulations. In a simulation with
a live main halo, it is hence possible that the $m=1$ torque from the main
halo may be weaker or even offset a portion of the torque from subhaloes,
therefore leading to weaker $m=1$ spiral structures. Calculating this effect
precisely requires dedicated, very high resolution simulations with a live dark
matter halo and is beyond the scope of this work.

\section{Removing subhalo A and B}
\label{sec:abc}

\begin{figure}
  \centering
  \includegraphics[width=\linewidth]{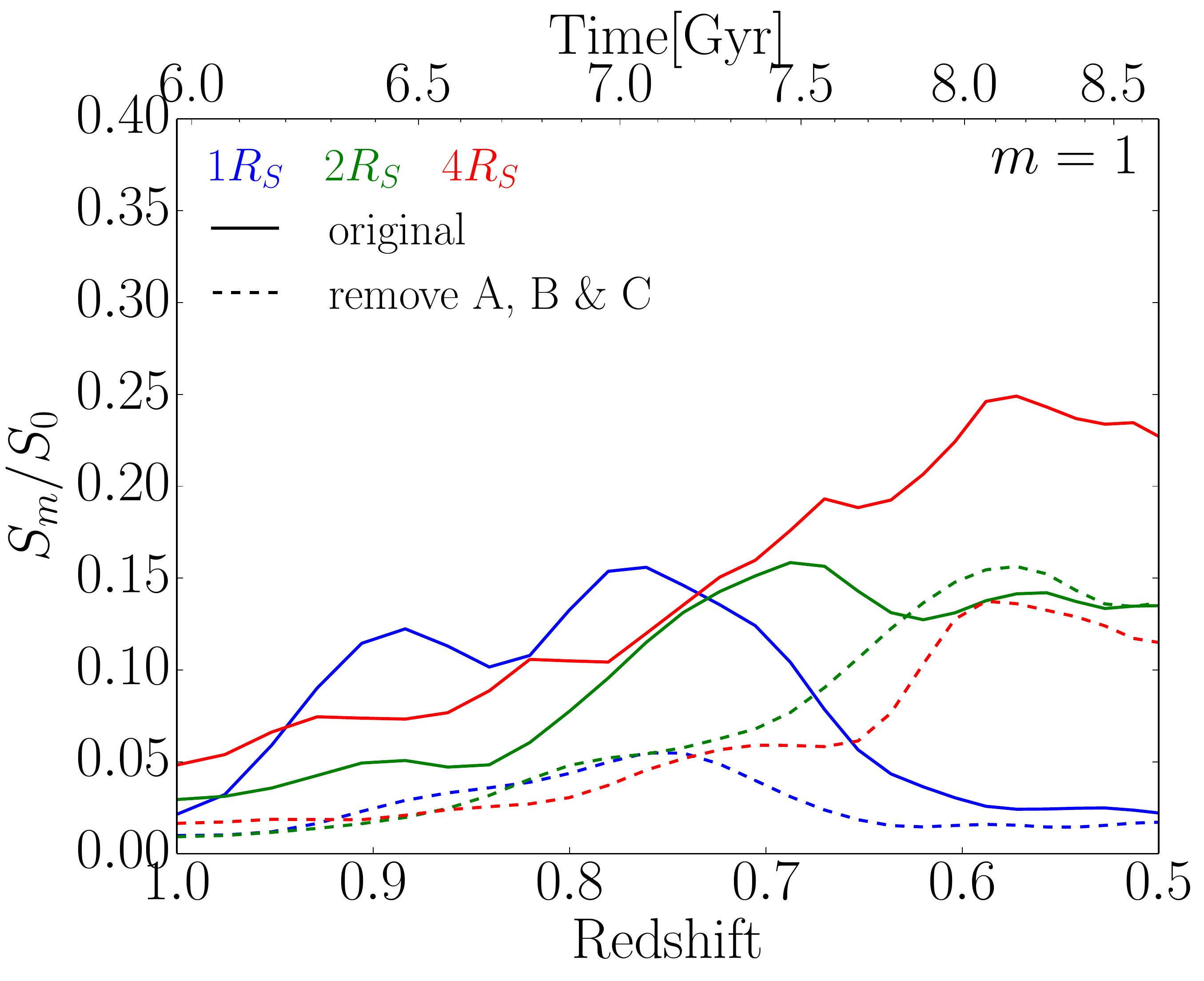}
  \caption{Simulation with subhaloes A and B removed. The first and the
    second generation of modes are absent, demonstrating clearly that
    these modes are mostly caused by subhaloes A and B.} 
    \label{fig:82}
\end{figure}

Further to Section~\ref{sec:eimp} where we study the impact of each halo, we
also run a simulation with subhalo A and B removed at a same time.
Note that since subhaloes A and C are the same subhalo hitting the disc twice,
in this simulation subhalo C is also removed. As shown in Figure~\ref{fig:82},
when subhaloes A and B are both removed, only very weak modes develop for $z >
0.7$, demonstrating that the first two generation of modes are mainly caused
by subhaloes A and B.

\section{Subhalo inspirals in live and static haloes}
\label{sec:livevsstatic}
\begin{figure*}
  \centering
  \includegraphics[width=\linewidth]{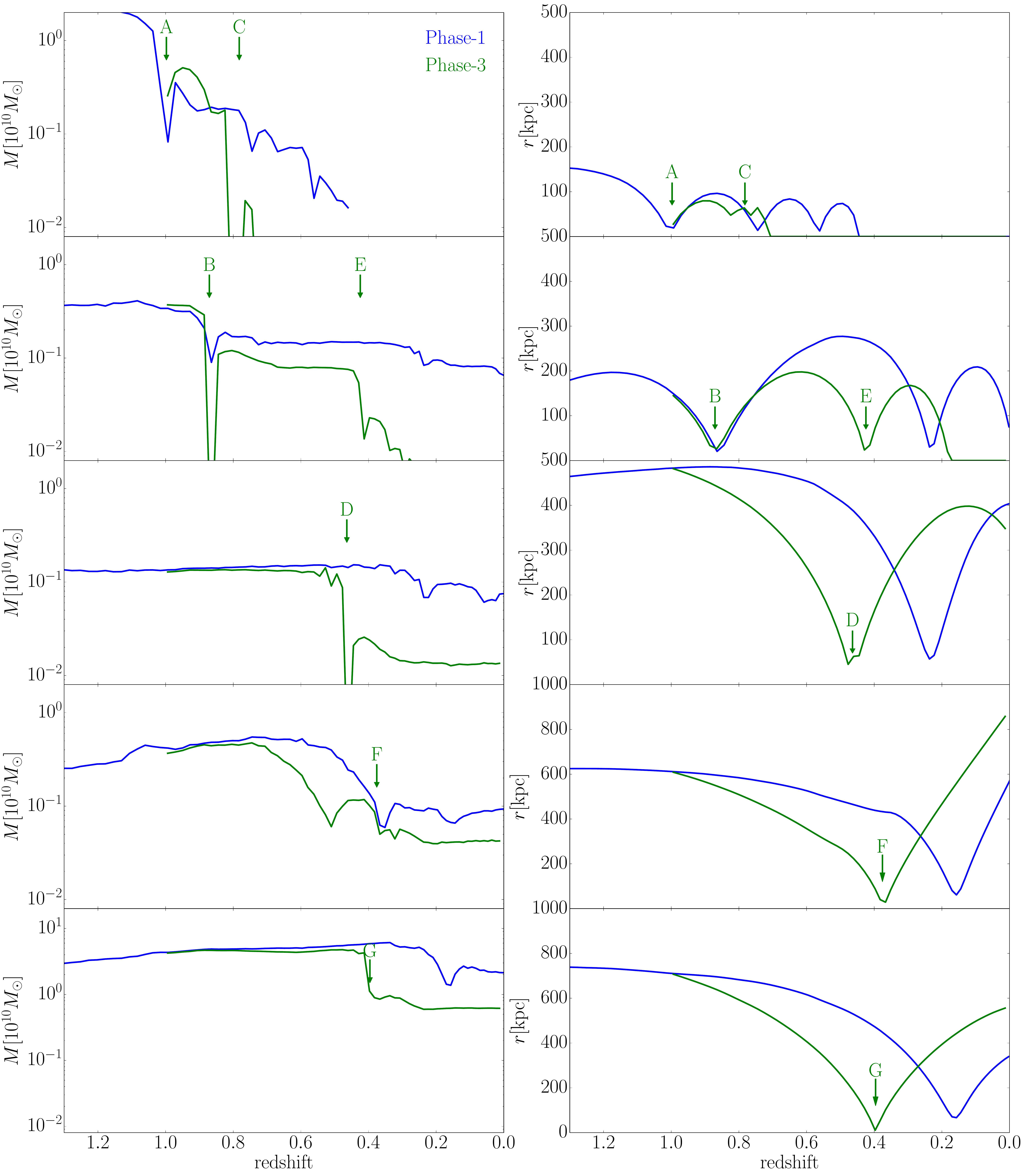}
  \caption{Mass evolution (left) and distance from the centre (right) of
    subhaloes as they pass through the central region of the halo and the
    disc. The impact times in the Phase-3 simulation of the subhaloes are
    marked with vertical arrows. Up to the passage through the pericentre mass
    loss is comparable in the two simulations. For earlier impacts (subhaloes
    A, B and C), the impacting time corresponds very well between the two
    simulations, while impacts happen earlier in the Phase-3 simulation for
    subhaloes D to G. Note that subhaloes A/C and B/E are largely disrupted at
    the later times of the simulation, where we truncate the curves.}
  \label{fig:subhalo-reduction}
\end{figure*}

To study the influence of substituting the live main dark matter halo with an
analytic halo potential, we compare the inspiral of
subhaloes listed in Table~\ref{tab:1} in the Phase-1 and Phase-3
simulation. The evolution of each subhalo's mass and distance from the centre
of the main halo is illustrated in Figure~\ref{fig:subhalo-reduction}. As
shown in left panels, subhaloes lose significant amount of mass every time
they go through the central region of the main halo. The mass loss with a live
halo (in blue) and with an analytic halo (in green) is generally comparable up
to the pericentric passage, ranging from $50\%$ to $80\%$. However, after the
passage trough the stellar disc the mass loss of subhaloes is typically larger
in the Phase-3 than in the Phase-1 simulation (which does not contain a live
stellar disc).     

As can be seen from the panels on the right, for subhaloes A, B and C the time
of the pericentric passage in the two simulations is very similar (as majority
of their trajectory has been computed prior to the start of the Phase-3 with
the live dark matter halo present), while other subhaloes reach pericentre
somewhat earlier in the Phase-3 than in the Phase-1 simulation. This is due to
the fact that in the Phase-3 simulation, the lack of the main dark matter halo
leads to a much lower dynamical friction on the subhaloes. Nonetheless, this
should not affect our results regarding the response of the stellar disc to the
subhaloes. In fact, we have verified that the velocity of the impact is
comparable (within a factor of $1.2$) and that the time subhaloes spend within
half of the apocentre is comparable as well (within factors of $0.85$
to $1.17$).

\end{document}